\documentclass[%
reprint,
aps, 
prstab, 
nofootinbib, 
longbibliography, 
floatfix, 
superscriptaddress,
]{revtex4-2}

\usepackage[utf8]{inputenc}

\usepackage{graphicx}

\usepackage[caption=false]{subfig}

\usepackage{xcolor}

\definecolor{colorq}{RGB}{15,92,171}
\definecolor{colors}{RGB}{237,176,45}
\definecolor{colort}{RGB}{179,29,63}
\newcommand*{\xq}{\color{colorq}} 
\newcommand*{\xs}{\color{colors!65!colort}} 
\newcommand*{\xt}{\color{colort}} 

\usepackage{amsmath,amssymb,textcomp}

\usepackage[per-mode=symbol]{siunitx}
\DeclareSIUnit{\foot}{ft}
\DeclareSIUnit{\voltampere}{VA}
\DeclareSIUnit{\sideg}{deg}
\DeclareSIUnit{\atm}{atm}

\newcommand*{\T}{\mathrm{T}}
\renewcommand*{\vec}[1]{\boldsymbol{\mathrm{#1}}}
\DeclareMathOperator{\diag}{diag}

\let\origim\Im
\renewcommand{\Im}[1]{\origim\left\lbrace#1\right\rbrace}

\newcommand{\underbracex}[2]{%
	\,\begin{array}[t]{@{}c@{}}
		\underbrace{#1}\\
		#2
	\end{array}\,
}

\usepackage{todonotes}
\let\origtodo\todo
\renewcommand{\todo}[1]{{\origtodo[inline]{#1}}}

\begin{document}



\title{rf measurements and tuning of the 1-meter-long 750~MHz radio-frequency quadrupole for artwork analysis}

\newcommand*{\affCERN}{\affiliation{European Organization for Nuclear Research (CERN), CH-1211 Geneva 23, Switzerland}}
\newcommand*{\affUni}{\affiliation{Institute of General Electrical Engineering, University of Rostock, D-18051 Rostock, Germany}}
\newcommand*{\affLLM}{\affiliation{Department Life, Light \& Matter, University of Rostock, D-18051 Rostock, Germany}}

\author{Hermann W. Pommerenke}%
\email{hermann.winrich.pommerenke@cern.ch}%
\affCERN
\affUni

\author{Ursula van Rienen}
\affUni
\affLLM

\author{Alexej Grudiev}
\affCERN

\date{\today}


\begin{abstract}
The 750~MHz PIXE-RFQ (radio-frequency quadrupole), developed and built at CERN, provides 2~MeV protons over a length of one meter for proton-induced X-ray emission analysis (PIXE) of cultural heritage artwork. In this paper, we report low-power rf measurements and tuning of the PIXE-RFQ, which have been completed mid-2020. Using a novel algorithm based on direct measurements of the response matrix, field and frequency could tuned at the same time in only two steps to satisfying accuracy. Additionally, we report measurements of single modules, quality factors, and coupling strength. In all cases, very good agreement between rf measurement and design values was observed.
\end{abstract}

\maketitle


\section{Introduction}

\begin{figure}[tbh]
	\centering
	\includegraphics[width=\linewidth]{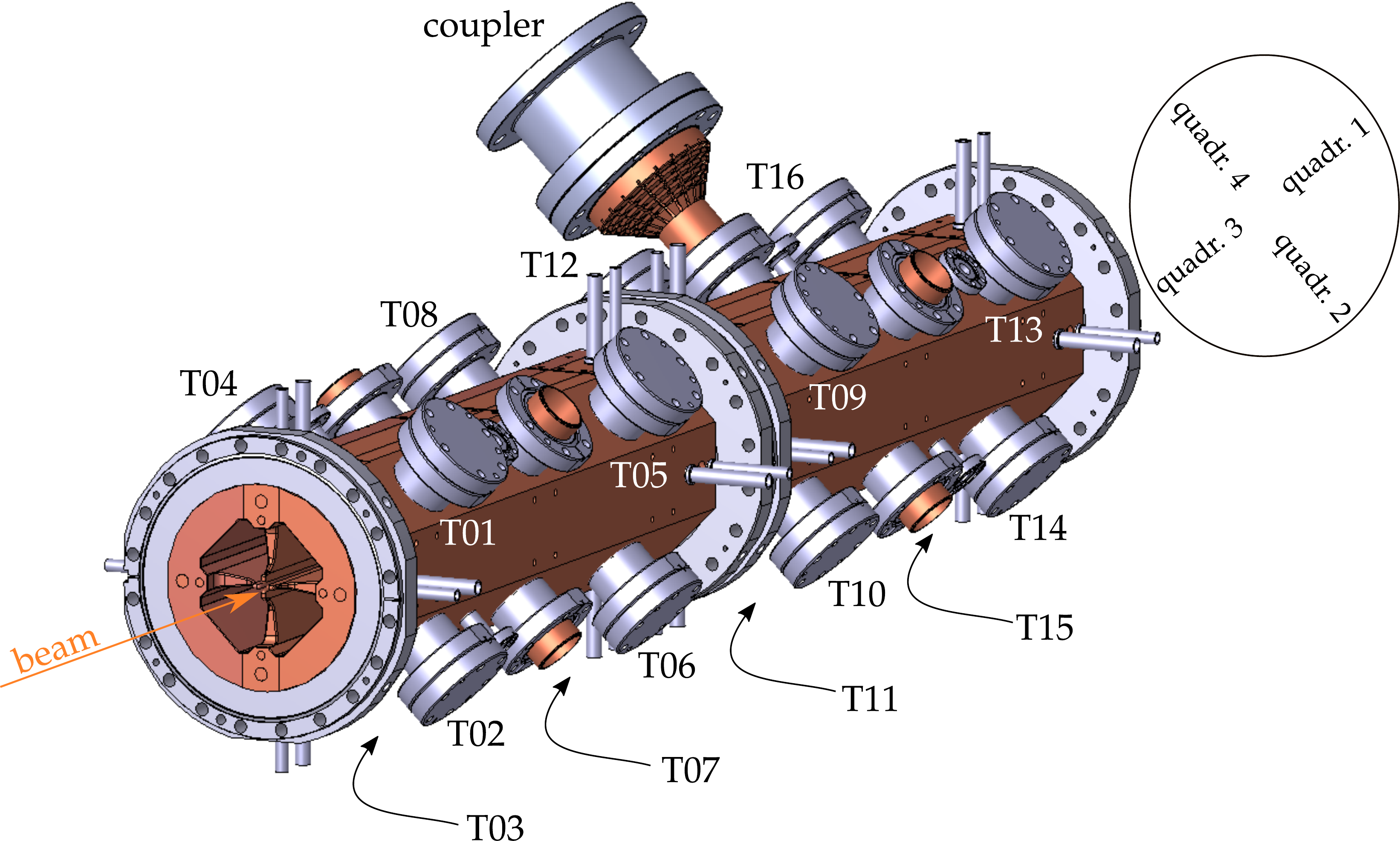}
	\caption{CAD model of the PIXE-RFQ (without end plates) and labeling of the quadrants and slug tuners. The arrows indicate the labels of hidden tuners in quadrant~3.}
	\label{fig:tuner_labels}
\end{figure}

After successful design, construction, and commissioning of the HF-RFQ, a compact \SI{750}{\mega\hertz}  radio-frequency quadrupole~(RFQ) for medical applications~\cite{Vretenar2014compact, Lombardi2015Beam, Vretenar2016High, Koubek2016Tuning, Koubek2017rf, Koubek2017rfrep, Dimov2018Beam, Lombardi2018High}, CERN initiated the development of a new RFQ operating at this high frequency in mid-2017. The so-called PIXE-RFQ~\cite{Vretenar2016High, Lombardi2018High, Pommerenke2018RF, Pommerenke2019rf, Mathot2019CERN} will accelerate protons to \SI{2}{\mega\eV} over a length of one meter only. A CAD model is shown in Fig.~\ref{fig:tuner_labels} and the RFQ key parameters are listed in Table~\ref{tab:key_parameters}.

\begin{table}[tbh]
	\centering
	\caption{Design parameters of the PIXE-RFQ~\cite{Lombardi2015Beam,Vretenar2016High}.}
	\begin{tabular}{l@{\qquad}rl}
			\hline\hline
			Species & proton & (H\textsuperscript{+}) \\
			Input energy & 20 & \si{\kilo\eV} \\
			Output energy & 2 & \si{\mega\eV} \\
			rf frequency & 749.480 & \si{\mega\hertz}\\
			Inter-vane voltage & 35 & \si{\kilo\volt} \\
			RFQ length & 1072.938 & \si{\milli\meter} \\
			Vane tip transverse radius & 1.439 & \si{\milli\meter} \\
			Mid-cell aperture & 1.439 & \si{\milli\meter} \\
			Minimum aperture & 0.706 & \si{\milli\meter} \\
			Final synchronous phase & -15 & \si{\sideg}\\
			Output beam diameter & 0.5 & \si{\milli\metre} \\
			Beam transmission & 30 & \si{\percent} \\
			Peak beam current  & 200 & \si{\nano\ampere} \\
			Repetition rate & 200 & \si{\hertz} \\
			Pulse length & 125 & \si{\micro\second} \\
			Duty factor & 2.5 & \si{\percent}\\
			Unloaded quality factor ($Q_0$) & 6000 & \\
			Peak rf power loss & 65 & \si{\kilo\watt} \\
			rf wall plug power & $\leq 6$ & \si{\kilo\voltampere}\\
			\hline\hline
	\end{tabular}
	\label{tab:key_parameters}
\end{table}

The PIXE-RFQ has been developed in the context of the MACHINA collaboration (Movable Accelerator for Cultural Heritage In-situ Non-destructive Analysis) between CERN and INFN~\cite{Giuntini2018MACHINA, Mathot2019CERN}. The aim of the project is to build the first transportable system for proton-induced X-ray emission analysis~(PIXE) of cultural heritage artwork, allowing employment in museums, restoration centers, or even in the field. Low-power rf measurements and tuning of the \SI{750}{\mega\hertz} four-vane RFQ have been completed in mid-2020.

In any four-vane RFQ, field tuning plays an important role to achieve the desired transverse and longitudinal field distribution. Vane modulation, certain design choices (e.g. a piecewise-constant cross section, as implemented here~\cite{Pommerenke2019rf}), as well as manufacturing imperfections and misalignments lead to local variations of the capacitance and inductance distribution.
Consequently, the ideal quadrupole field of the TE\textsubscript{210} operating mode is perturbed~\cite{Balleyguier20003D,Wangler2008RF}. The perturbation must be corrected by means of bead-pull measurements and movable piston tuners that allow for locally modifying the inductance. Many rf cavities and RFQs in particular are tuned using transmission line models (see e.g. Refs.~\cite{France2002Theoretical, Palmieri2010Perturbation, Tan2014Simple, Rossi2012Assembly, Piquet2013RF}). Contrarily, we adopted the tuning algorithm developed for the HF-RFQ by Koubek~et~al.~\cite{Koubek2017rf, Koubek2017rfrep} based on Ref.~\cite{Wangler2008RF}, where the effects of idividual tuner movements on the field are directly measured and recorded in a response matrix. Corrective tuner movements are than obtained by matrix inversion. We augmented this algorithm such that both frequency and field could be tuned at the same time.

This paper covers low-power rf measurements and tuning procedure of the PIXE-RFQ. In the following, after reviewing preliminary considerations to the measurements (Section~\ref{sec:prerequisites}), we report the single module measurements carried out on the individual mechanical modules in Section~\ref{sec:single_module}. Section~\ref{sec:tuning} discusses both algorithm and execution of the tuning procedure. Lastly, quality factor measurements are reported in Section~\ref{sec:Q}.


\section{Preliminary considerations}
\label{sec:prerequisites}

This section briefly describes the bead-pull measurement, lists the tuning goals, and discusses corrections of frequency measurements with respect to temperature and medium.





\subsection{Bead-pull field measurement}

In bead-pull measurement, a small object (the bead) is introduced into the rf cavity, effectively removing a small volume from the resonator. Following Slater's perturbation theorem~\cite{Slater1946Microwave, Maier1952Field} this can be observed as a change in resonance frequency of the cavity, proportional to the squared field amplitudes at the location of the bead. In RFQs the bead is typically introduced into the four quadrants, i.e. the regions occupied by the magnetic field.

In the present case, the bead-pull measurements were performed using the phase \mbox{$\Delta\phi=\arg S_{11}$} of the reflection \mbox{$S_{11}=\Gamma$} measured through the single input power coupler (upper part of Fig.~\ref{fig:tuner_labels}). The same bead-pull bench and pulley system as for the HF-RFQ was used~\cite{Koubek2017rfrep, Koubek2017rf}. The bead was threaded on a \SI{0.3}{\milli\meter} fishing wire. The size of the perturbing bead must be small enough such that the frequency shift stays within the linear regime of the $\Delta\phi(\Delta f)$~curve, but large enough such that an acceptable signal-to-noise ratio~(SNR) is achieved. Following the measurements of the HF-RFQ, a $\SI{7}{\milli\meter}\times\SI{4}{\milli\meter}$ aluminum bead was used. While the PIXE-RFQ features the same frequency and a comparable quality factor, its length and thus volume are approximately halved. Therefore, the same bead introduces roughly double the frequency shift when inserted into the PIXE-RFQ. With $\Delta f\approxeq\SI{5}{\kilo\hertz}$, or $\Delta\phi\approxeq\SI{17}{\degree}$, the shift induced by the bead still remained well within the linear regime. 

From the raw phase measurements of the RFQ quadrants the field components were extracted and aligned by means of a few data processing steps detailed in Ref.~\cite{Koubek2017rfrep} and a smoothing Savitzky-Golay filter~\cite{Savitzky1964Smoothing, Press2007Numerical}. With $\Delta\phi$ proportional to the squared magnetic field, the relative quadrant amplitudes \mbox{$a_1,\dots, a_4$} arise by taking the square root and assigning the proper sign to account for the the alternating field orientation of the TE\textsubscript{210} mode. The field flatness is then quantified by one quadrupole two dipole components of orthogonal polarizations~\cite{Wangler2008RF}:
\begin{equation}
\begin{split}
Q\phantom{^\text{S}} &= \frac{1}{4}\left(a_1 - a_2 + a_3 - a_4\right),\\
D^\text{S} &= \frac{1}{2}\left(a_1 - a_3\right),\\
D^\text{T} &= \frac{1}{2}\left(a_2 - a_4\right).
\end{split}\label{eq:bpull_quad_dip_components}
\end{equation}


\subsection{Tuning goals}
\label{sec:tuning_goals}

By design, the inter-vane voltage of the PIXE-RFQ is constant at $V_0=\SI{35}{\kilo\volt}$, corresponding to a magnetic field that is constant along the RFQ and equal in all four quadrants. The goal of the tuning process can thus be defined as \mbox{$Q=\SI{100}{\percent}=\text{const.}$} and \mbox{$D^\text{S}=D^\text{T}=0$} at all sampling points. Errors of $\pm\SI{2}{\percent}$ with respect to the average quadrupole component are acceptable in each of the three field components from a beam dynamics point of view. These tolerances have been established with respect to the \SI{2}{\meter} long \SI{750}{\mega\hertz} HF-RFQ~\cite{Koubek2017rf}.

The PIXE-RFQ represents a stand-alone machine; no rf structures requiring frequency and phase stability are installed downstream of the RFQ. Therefore, frequency accuracy does not represent a critical tuning goal. A frequency shift is in fact foreseen by design during nominal operation in order to maintain a constant cooling water temperature~\cite{Pommerenke2019rf}. Deviations from the design resonant frequency as much as few \si{\mega\hertz} are acceptable for the beam dynamics~\cite{Pommerenke2019rf}. Nevertheless, the frequency was tuned to match the design value: \SI{749.480}{\mega\hertz} under vacuum at \SI{22}{\celsius} with an error smaller than~$\pm\SI{60}{\kilo\hertz}$. This tolerance emerges from the water temperature range of a typical cavity cooling system: $\pm\SI{5}{\kelvin}$ around the nominal value. The corresponding sensitivity of the PIXE-RFQ resonant frequency amounts to $-\SI{13.3}{\kilo\hertz\per\kelvin}$, as determined in Ref.~\cite{Pommerenke2019rf}.


\subsection{Frequency correction}

Most of the measurements were conducted under air since the bead-pull setup denies evacuating the RFQ cavity. Furthermore, the ambient temperature was not controlled and deviated significantly from the RFQ design reference temperature of \SI{22}{\celsius}. The measured frequency was therefore affected by two main aspects: (i)~thermal expansion of the RFQ cavity, and (ii)~the relative permittivity~$\varepsilon_\text{r}$ of air.

The thermal behavior of the cavity during low-level rf measurements was dominated by the ambient temperature that changed during the day. Because of thermal expansion, the resonant frequency is anti-proportional to the bulk copper temperature: \mbox{$\Delta f/f=-\alpha\Delta T$}, where $\alpha=\SI{1.66e-5}{\per\kelvin}$ is the secant thermal expansion coefficient of the copper cavity and $\Delta T$ the difference between measured and reference temperature. The $\pm\SI{0.2}{\kelvin}$~precision of the used thermometers translates to a frequency uncertainty of $\pm\SI{2.6}{\kilo\hertz}$. This error is by an order of magnitude smaller than the error introduced by humidity uncertainty (see the following).

The speed of light and thus the resonant frequency in a medium are by a factor of $\sqrt{\varepsilon_\text{r}\mu_\text{r}}$ lower than the vacuum values~\cite{Feynman2010Feynman}. For air we assume $\mu_\text{r}=1$ and $\varepsilon_\text{r}=1.00058986$~\cite{Hector1936dielectric} at standard temperature and pressure~(STP, \SI{0}{\celsius}, \SI{1}{\atm}, \SI{900}{\kilo\hertz}). The measured frequency~$f_\text{meas}$ is thus corrected as follows:
\begin{equation}
f = f_\text{meas}\cdot\frac{\sqrt{\varepsilon_\text{r}}}{1 - \alpha\Delta T}.\label{eq:freq_correction_T_air}
\end{equation}
A major source of uncertainty is the dependence of $\varepsilon_\text{r}$ on the ambient humidity. Analytical expressions for this relationship have been formulated based on experimental studies and can be found in Refs.~\cite{SantoZarnik2012experimental, Buck1981New, BRILLC2012Model}. The influence can be mitigated by measuring the humidity and calculating the corresponding~$\varepsilon_\text{r}$. Alternatively, the cavity can be flooded with dry nitrogen~(N\textsubscript{2}). The latter represents a standard procedure for rf cavity measurements and has been done during the final tuning steps of the medical HF-RFQ~\cite{Koubek2017rf}.

However, there are no strict accuracy requirements for the PIXE-RFQ frequency. Hence, we accepted the humidity uncertainty and worked with the constant STP value of $\varepsilon_\text{r}$. A humidity uncertainty of $\pm\SI{30}{\percent}$ translates to an error in frequency of approximately $\pm\SI{30}{\kilo\hertz}$. Additional, but much smaller errors originate in pressure and temperature dependence of $\varepsilon_\text{r}$~\cite{Heidary2010Low}. Thus, we expected that the PIXE-RFQ could be tuned under air to an accuracy of approximately $\pm\SI{30}{\kilo\hertz}$ when measuring only the cavity temperature. If frequency stability was required, this error would still lie within the $\pm\SI{5}{\kelvin}$ tuning range of a typical water cooling system.

In the following, the frequency is exclusively reported in terms of the corrected value (the value under vacuum at \SI{22}{\celsius}).


\section{Single module measurements}
\label{sec:single_module}

The PIXE-RFQ consists of two mechanical modules, in the following denoted as module~1 and module~2, which were brazed individually and then clamped together to form the full assembly. Both modules were measured individually to ascertain the manufacturing quality, and to determine if it would be necessary to use the vacuum pumping ports as ``emergency'' tuning features in addition to the piston tuners. By comparing both resonant frequency and field distribution to the simulation (eigenmode solver of CST Microwave Studio\textsuperscript{\textregistered}~2018~\cite{CST2018CST}) it was found that no special measures were required, as very close agreement was observed between measurement and simulation.

\begin{figure}[tbh]	
	\centering
    \includegraphics[width=\linewidth]{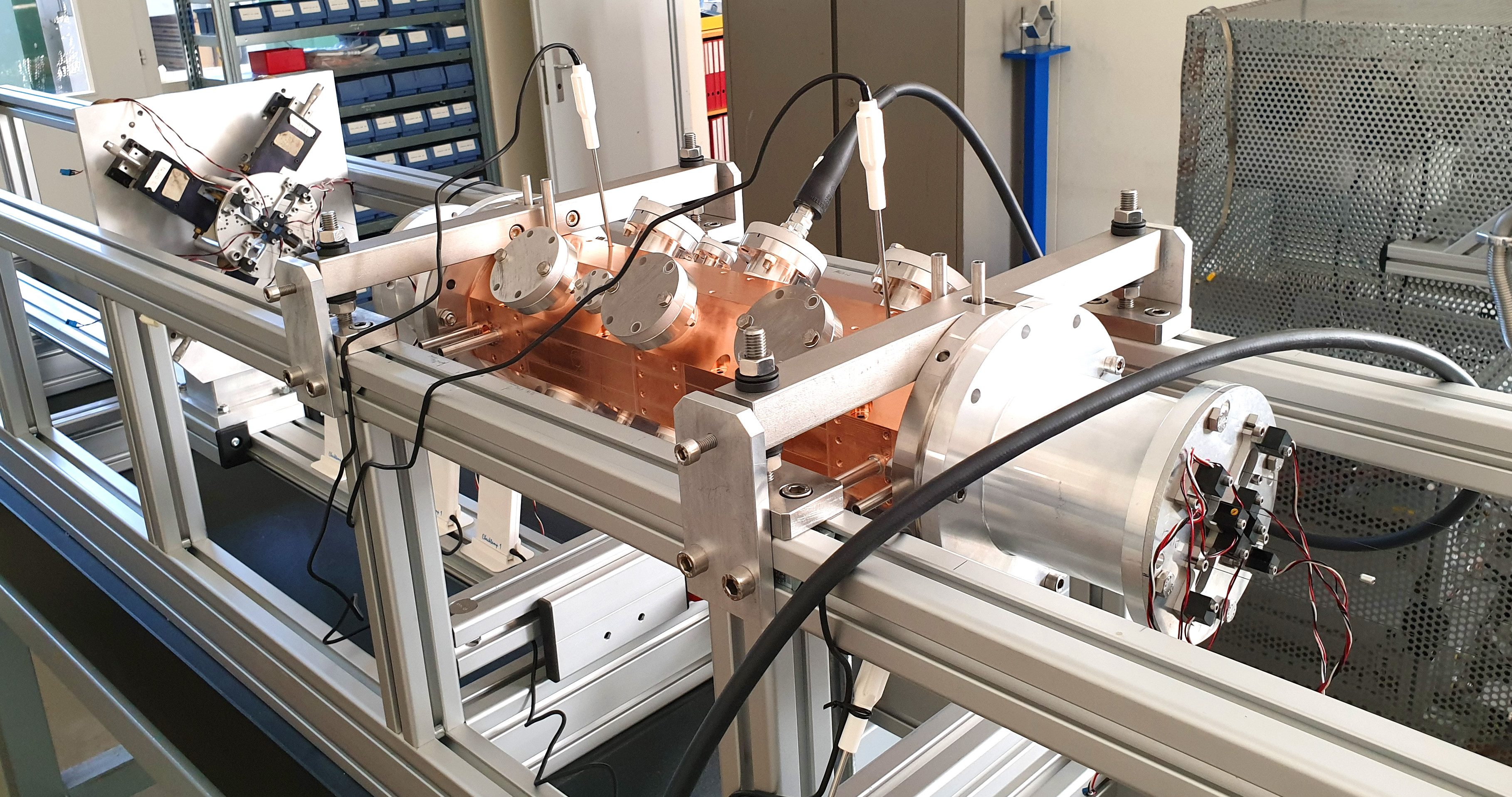}%
    \label{fig:photo_single_module_bpull}%
	\caption{Photograph of the first module of the PIXE-RFQ with extension tubes mounted on the bead-pull support frame for single module measurements.}
	\label{fig:photo_single_module}
\end{figure}

Figure~\ref{fig:photo_single_module_bpull} shows module~1 mounted to the measurement bench for the single module measurement. No auxiliaries were installed and all ports were closed by aluminum flange covers. In the absence of coupler or diagnostic pickup antennas, a small makeshift antenna crafted from simple copper wire was mounted to one of the ports. Although the antenna was strongly under-coupled, a phase shift of \SI{4}{\degree} was observed upon introducing the bead into the cavity. With an SNR of roughly \SI{45}{\dB} this was considered as sufficient. As no end plates were present, the upstream and downstream ends of the module were terminated by aluminum extension tubes in order to obtain well-defined boundary conditions that could be reproduced in a 3D eigenmode simulation. 

	

\begin{figure}[tbh]
	\centering
	
	\subfloat[module 1]{
	    \includegraphics[width=\linewidth]{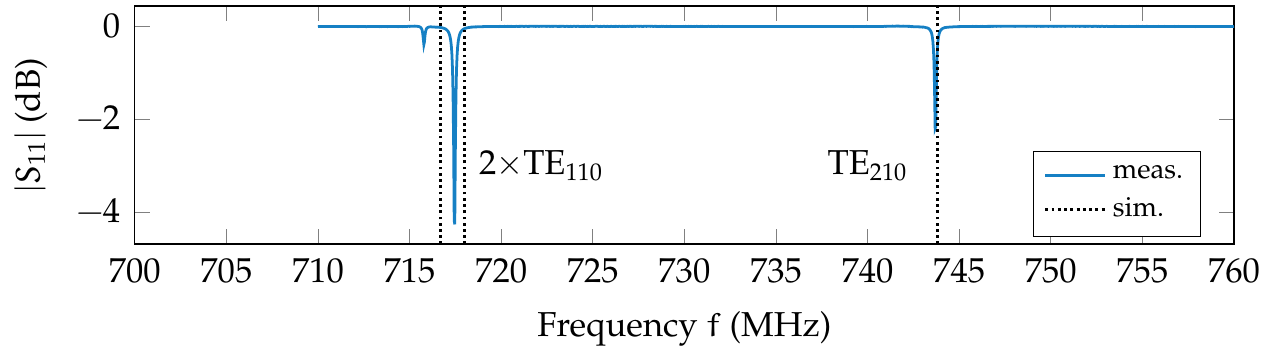}%
	    \label{fig:spectrum_single_module_1}%
	}
	
	\subfloat[module 2]{
	    \includegraphics[width=\linewidth]{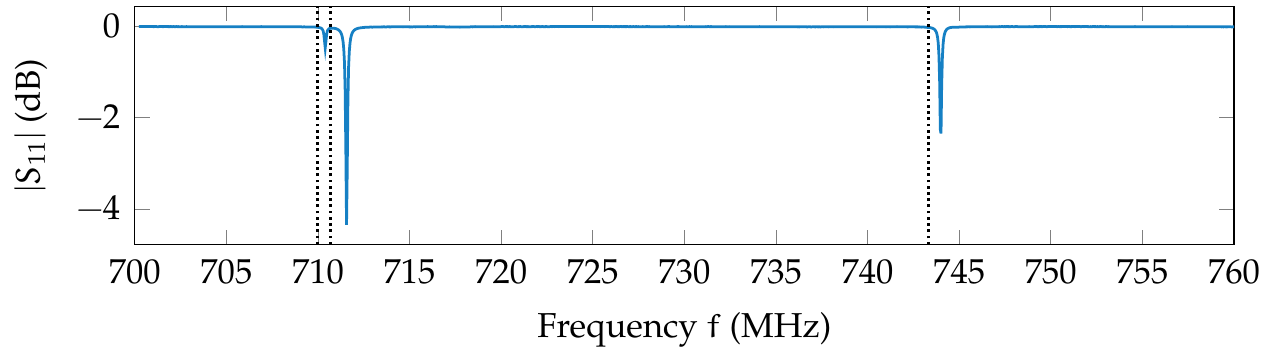}%
	    \label{fig:spectrum_single_module_2}%
	}
	
	\caption{Measured reflection coefficients of the individual RFQ modules using a small makeshift antenna. Because of this asymmetric arrangement, one of the dipole modes is excited much stronger than the other.}
	\label{fig:spectrum_single_module}
\end{figure}

Figures~\ref{fig:spectrum_single_module_1} and \subref{fig:spectrum_single_module_2} show the measured spectra of the individual modules with attached extension tubes. Because of the absence of any auxiliaries and the metallic extension tubes, the TE\textsubscript{210} frequencies are roughly \SI{6}{\mega\hertz} lower than the RFQ design frequency of \SI{749.48}{\mega\hertz}. The measured frequencies deviate from the simulation value by \SI{150}{\kilo\hertz} for module~1 and \SI{600}{\kilo\hertz} for module~2. The deviations in the dipole-mode frequencies are smaller than \SI{1}{\mega\hertz}.

\begin{figure}[tbh]
	\centering
	
	\subfloat[module 1]{
	    \includegraphics[width=\linewidth]{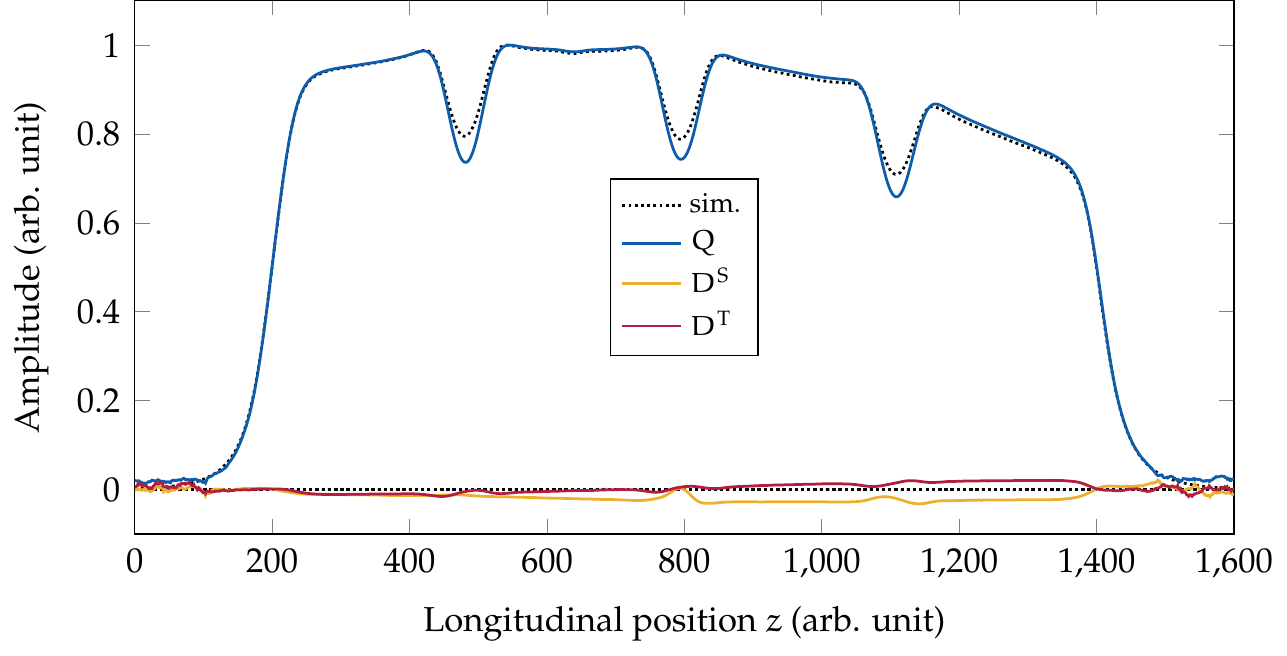}%
	    \label{fig:field_single_module_1}%
	}
	
	\subfloat[module 2]{
	    \includegraphics[width=\linewidth]{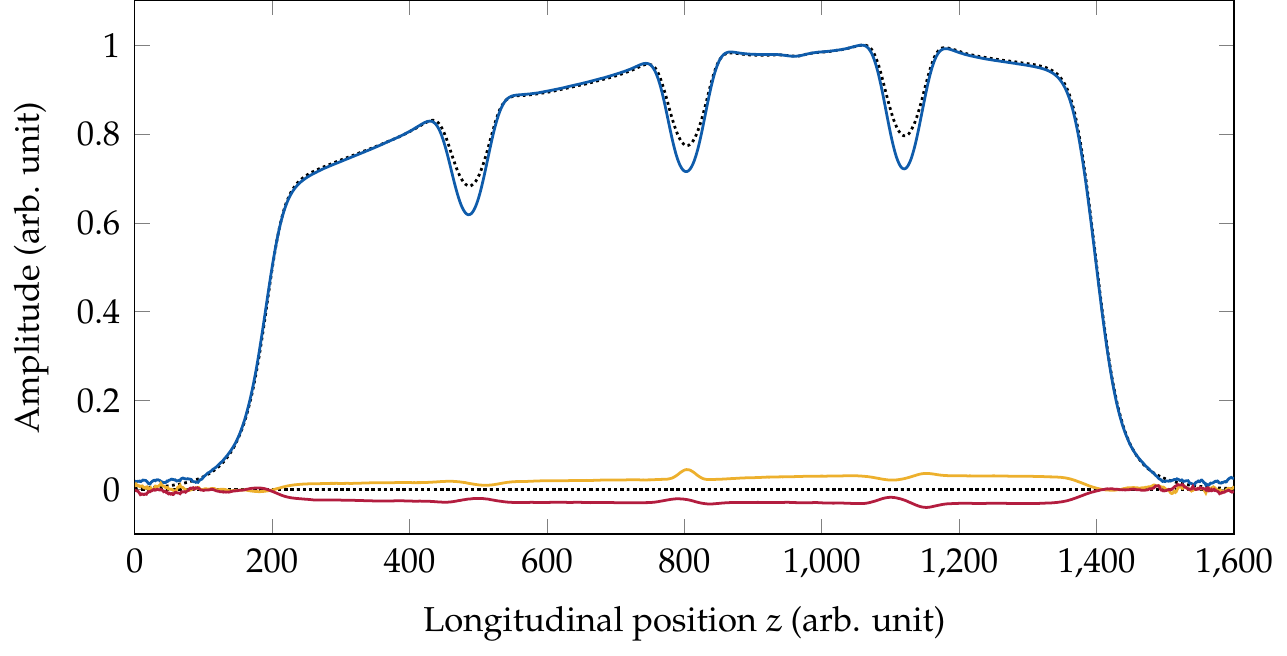}%
	    \label{fig:field_single_module_2}%
	}
	
	\caption{Measured quadrupole component~$Q$ and dipole components~$D^\text{S}$, $D^\text{T}$ of the individual RFQ modules in comparison to the simulation (dotted lines).}
	\label{fig:field_single_module}
\end{figure}

A near-perfect agreement between measured and simulated quadrupole component~$Q$ [see Eq.~\eqref{eq:bpull_quad_dip_components}] was observed with an error less than \SI{1}{\percent} of the average $Q$ amplitude~(Fig.~\ref{fig:field_single_module}). The measured dipole components~$D^\text{S}$, $D^\text{T}$, which vanish in the simulation, reach amplitudes of up to $\pm\SI{4}{\percent}$. Larger deviations were expected for the full assembly, as the clamped connection of the two modules introduces additional alignment errors.

\section{Tuning}
\label{sec:tuning}

After the assembly of the full structure---two modules, two end plates, pumping ports, power coupler, and diagnostic antennas---the PIXE-RFQ was tuned by means of sixteen movable piston tuners. In this section, the tuner tooling is described, reliability measurements are shown, the augmented tuning algorithm is derived, and the tuning steps are reported. Field and frequency after tuning and final tuner installation are compared to the initial values.


\subsection{Tuner setup and tooling}

\begin{figure}[tbh]	
	\centering
	
	\subfloat[]{
	    \includegraphics[width=0.55\linewidth]{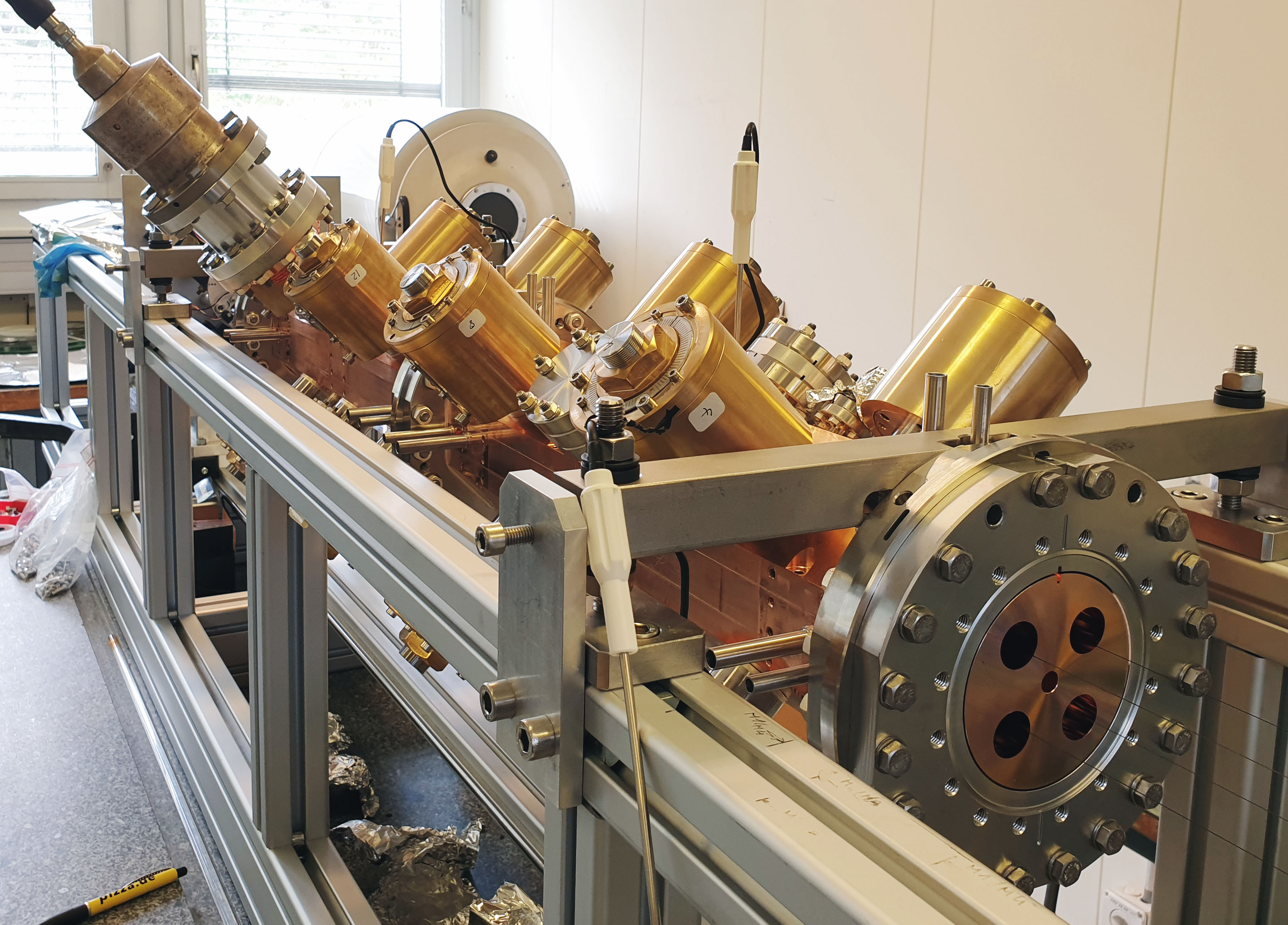}%
	    \label{fig:photo_full_assembly}%
	}%
	\hfill%
	\subfloat[]{
	    \includegraphics[width=0.421\linewidth]{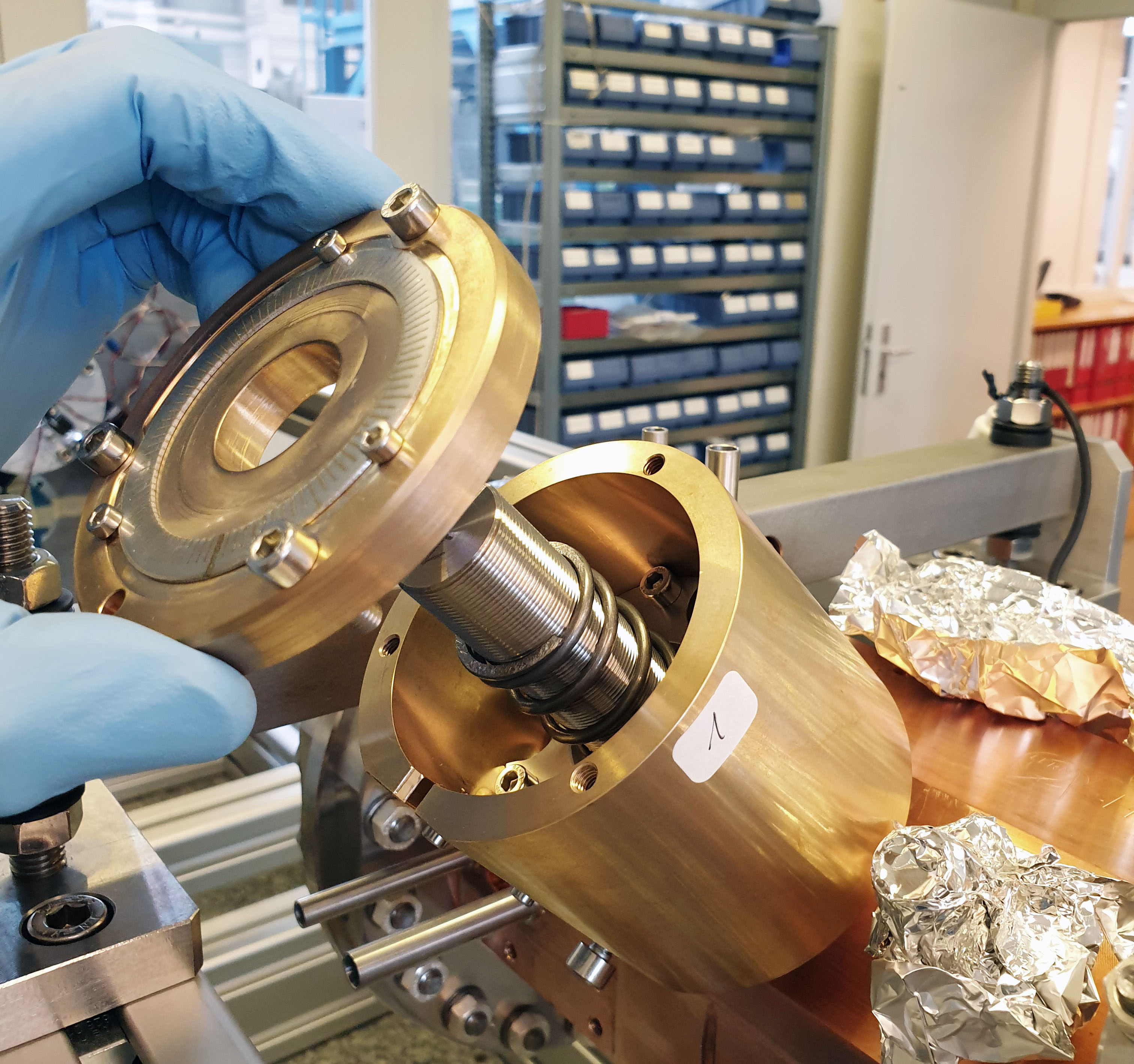}%
	    \label{fig:photo_tuner_tooling_assembly}%
	}
	
	\subfloat[]{
	    \includegraphics[width=0.48\linewidth]{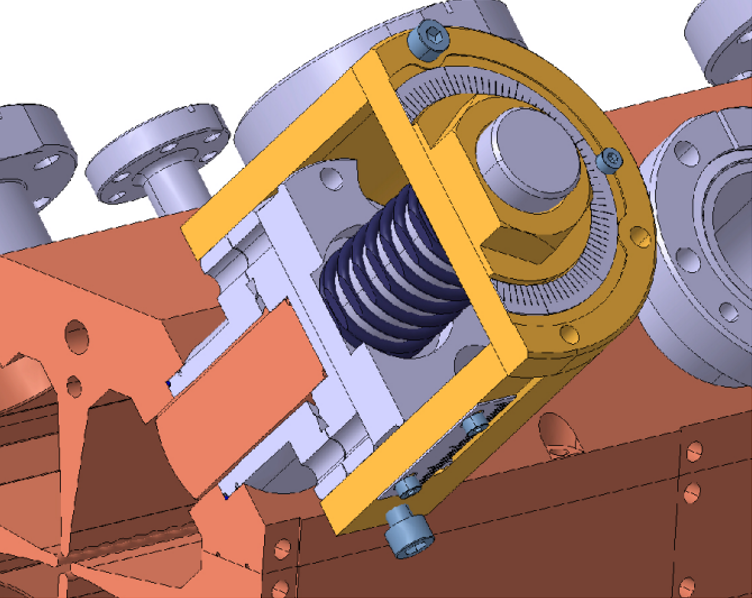}%
	    \label{fig:tuner_tooling_cad}%
	}%
	\hfill%
	\subfloat[]{
	    \includegraphics[width=0.48\linewidth]{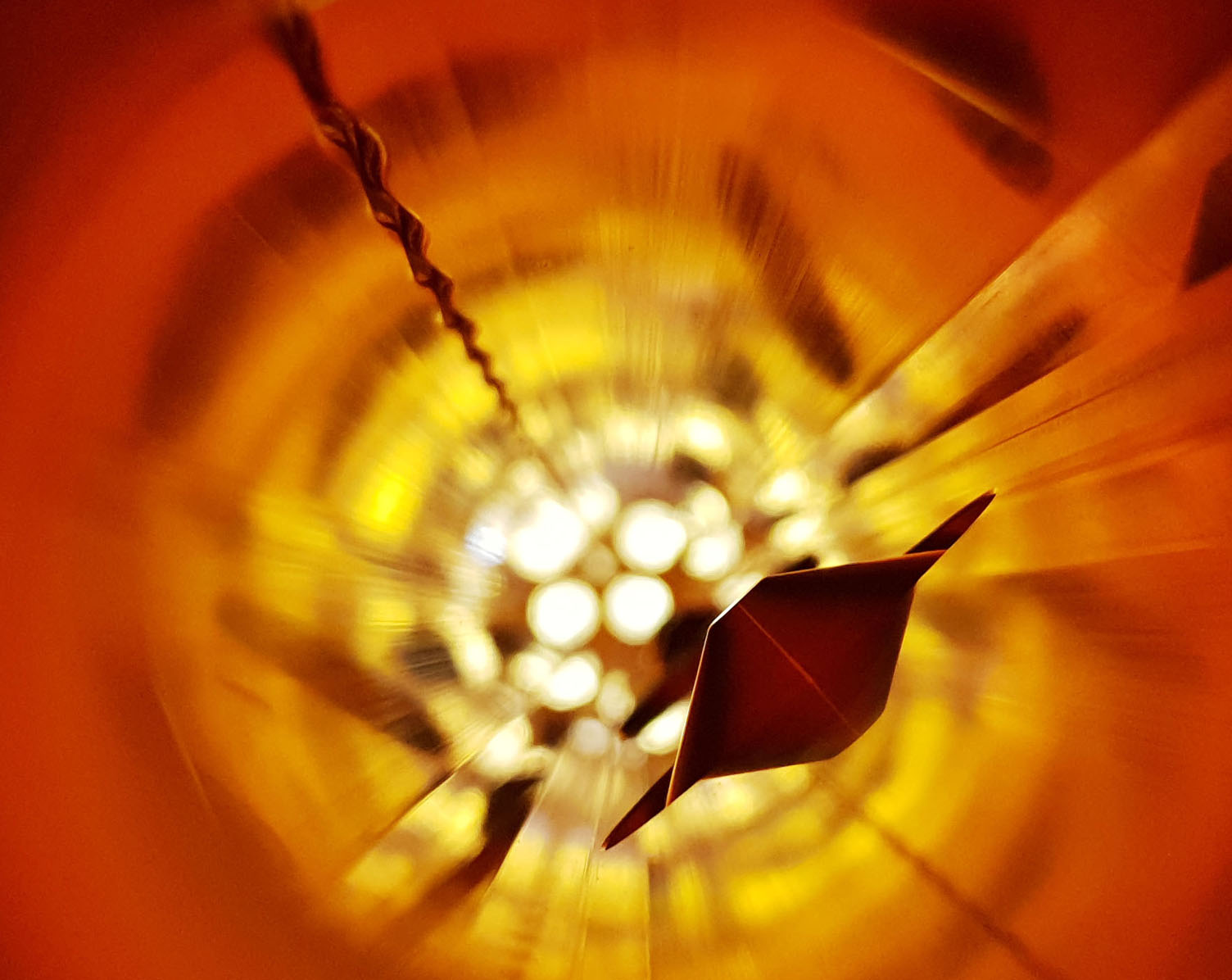}%
	    \label{fig:photo_tuner_in_cavity}%
	}
	
	\caption{PIXE-RFQ on bead-pull bench with tuners mounted in guidance tubes~\protect\subref{fig:photo_full_assembly}. Closeup of tuner tooling is shown in~\protect\subref{fig:photo_tuner_tooling_assembly}. CAD model of the tuner tooling is shown in~\protect\subref{fig:tuner_tooling_cad}; taken from Ref.~\cite{Koubek2017rf} with courtesy of B.~Koubek. Tip of tuner inside RFQ cavity seen from one of the bead-pull holes is shown in~\protect\subref{fig:photo_tuner_in_cavity}.}
	\label{fig:photo_tuning_setup}
\end{figure}

The PIXE-RFQ features sixteen piston tuners, copper slugs with a conical tip [Fig.~\ref{fig:photo_tuner_in_cavity}]. Four tuners each are mounted at four positions along the RFQ. Their designations are shown in Fig.~\ref{fig:tuner_labels}.

	
	

The nominal tuner length is given by the nominal tuner insertion into a perfect cavity while mounted flush in the so-called flange-to-flange position. Initially the tuners were machined with an over-length of \SI{11}{\milli\meter}, allowing for a mechanical tuning range of~$\pm\SI{11}{\milli\meter}$. A final length for each tuner was determined by iterative tuner adjustments and bead-pull measurements. The tooling was originally developed for the HF-RFQ~\cite{Koubek2017rf,Koubek2017rfrep} and is shown in Figs.~\ref{fig:photo_tuner_tooling_assembly} and~\subref{fig:tuner_tooling_cad}. During the tuning, the slugs were mounted on a threaded, spring-loaded piston, allowing for adjustment of the tuner penetration into the RFQ. The piston was mounted in a guidance tube fixed to the RFQ flange, which offered accurate transverse positioning. A scale with \SI{10}{\micro\meter} graduation was used to adjust the tuner position. The penetration was confirmed by means of a caliper before the tooling was removed after the final tuning step. Then, the tuners were remachined to their final lengths and installed flange-to-flange with copper gaskets.

\begin{figure}[tbh]
	\centering
	\includegraphics[width=\linewidth]{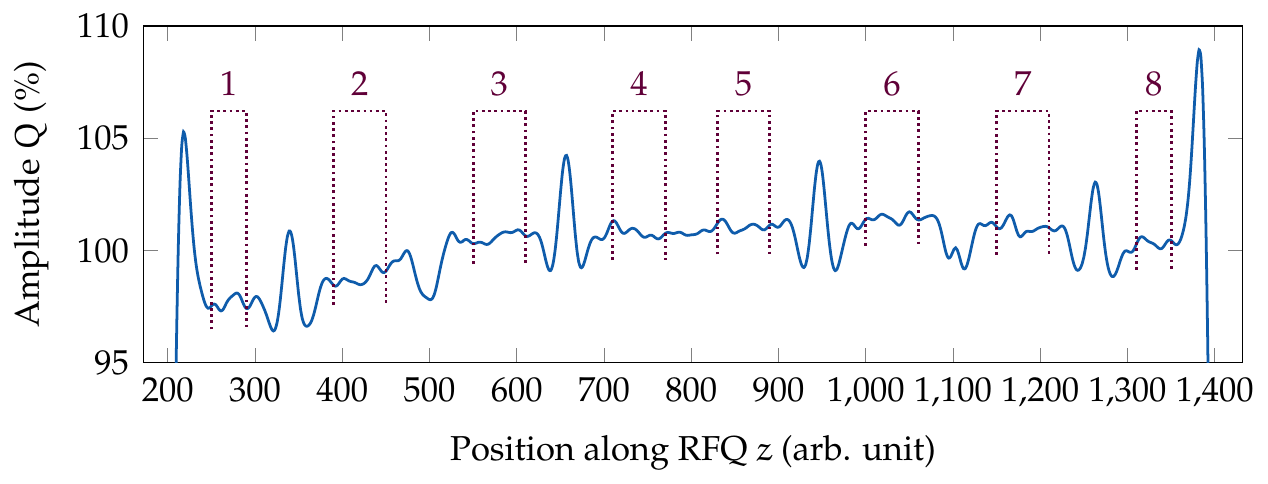}
	\caption{Example measurement of $Q$ component before tuning. Brackets visualize the windows over which the measurement is averaged to obtain the discrete sampling points.}
	\label{fig:field_sampling}
\end{figure}

During the tuning procedure, the field quality was assessed by measuring the quadrupole and dipole components of the TE\textsubscript{210} eigenmode, $Q$, $D^\text{S}$, and~$D^\text{T}$~[Eq.~\eqref{eq:bpull_quad_dip_components}] at discrete points. Each sampling point corresponds to an interval of the continuous field profile over which the data were averaged to reduce noise (Fig.~\ref{fig:field_sampling}).

\subsection{Reliability measurements}

Before the tuning process, the measurement error was estimated by means of reliability measurements. The authors differentiate between two distinct concepts: (i)~repeatability, the error observed between multiple bead-pull measurements taken without any changes to the RFQ itself, in particular no tuner movements, and (ii)~reproducibility, the error introduced by tuner movements and mechanical hysteresis effects. The errors arising from these two aspects pose the ultimate limit for the tuning accuracy.

\subsubsection{Repeatability of field measurement}

\begin{figure}[tbh]
	\centering
	\includegraphics[width=\linewidth]{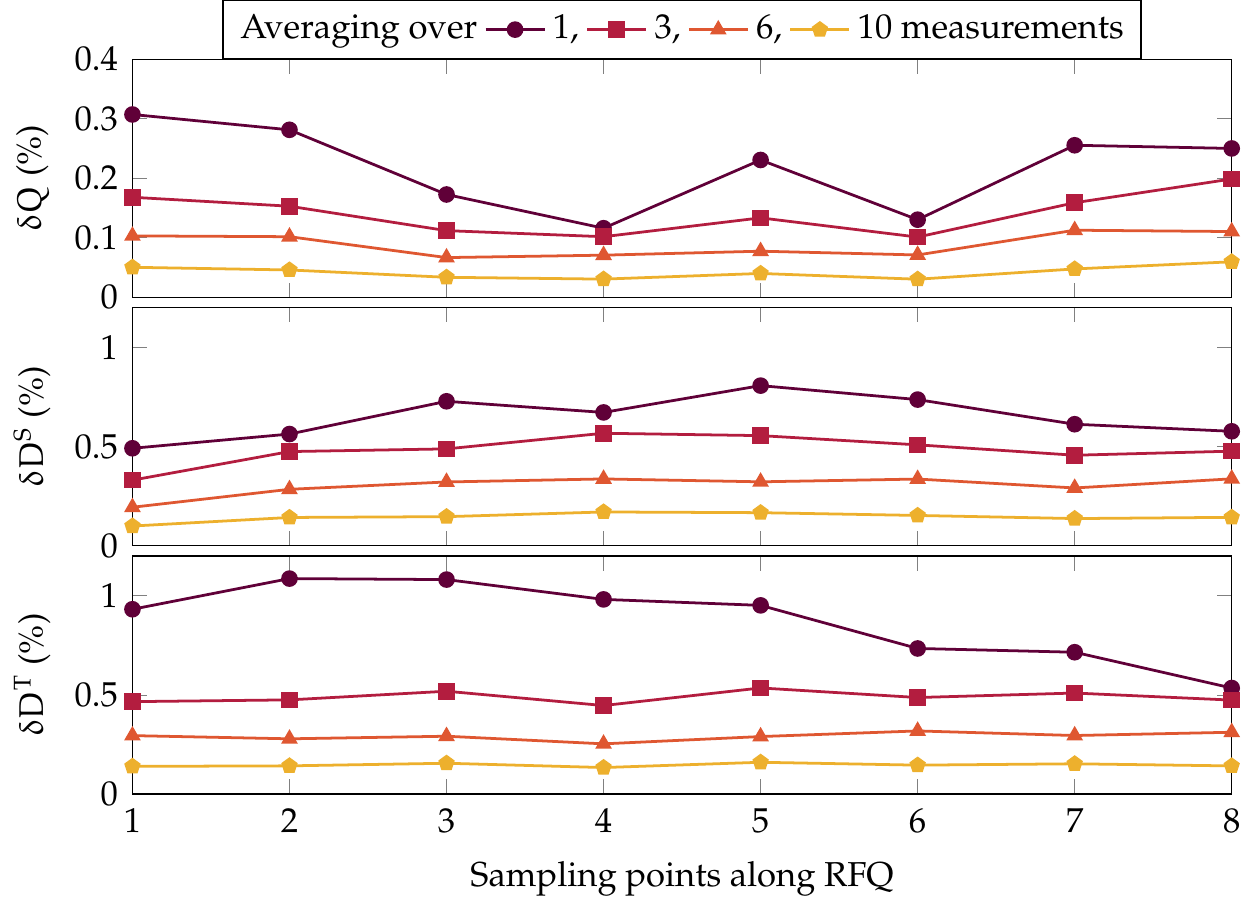}
	\caption{Reduction of error in quadrupole component ($\delta Q$) and dipole components ($\delta D^\text{S}$,~$\delta D^\text{T}$) when averaging over three, six, and ten measurements, compared to no averaging (one measurement). The source data was obtained from thirteen measurements carried out over several days.}
	\label{fig:field_error_averaging}
\end{figure}

To assess repeatability, thirteen bead-pull measurements were carried out over several days without moving any tuners. The deviations between the repeated measurements were larger than \SI{0.5}{\percent} for one quadrant. Such an error would limit the dipole-component tuning accuracy to more than \SI{1}{\percent} (Fig.~\ref{fig:field_error_averaging}). The deviations comprise systematic errors caused by change of ambient parameters such as temperature and humidity. However, they were expected to be negligible in this case, being several orders of magnitude smaller than what could be resolved by bead-pull measurement. A significantly larger error source is random noise introduced by vibration and slippage of the wire and by the vector network analyzer (VNA) itself.

Therefore, we studied to which extent the error could be reduced by averaging over several repetitions. Based on the data from the thirteen runs, all possible combinations of three, six, or ten measurements where formed. The average of each combination was calculated at each sampling point, generating new artificial sets of measurements. The resulting error was calculated as \mbox{$\delta X = \max \left| X - \left\langle X \right\rangle\right|$}, where \mbox{$X\in\{Q, D^\text{S}, D^\text{T}\}$} and $\langle\cdot\rangle$ indicates averaging over all thirteen available measurements. The results are summarized in Fig.~\ref{fig:field_error_averaging}, given as a percentage of the quadrupole component ($\left\langle Q\right\rangle=\SI{100}{\percent}$). Averaging over three measurements guarantees a measurement error $<\SI{0.5}{\percent}$ for $D^\text{S},D^\text{T}$ and $<\SI{0.2}{\percent}$ for $Q$, which lies well within the tuning requirements. Especially for sampling points with large spread, the $Q$ error could be reduced by nearly a factor of two. With respect to the considerable effort for carrying out these measurements, it was decided to repeat all tuning-related bead-pull measurements three times.

\subsubsection{Reproducibility with respect to tuner position}

\begin{figure}[tbh]
	\centering
	\includegraphics[width=\linewidth]{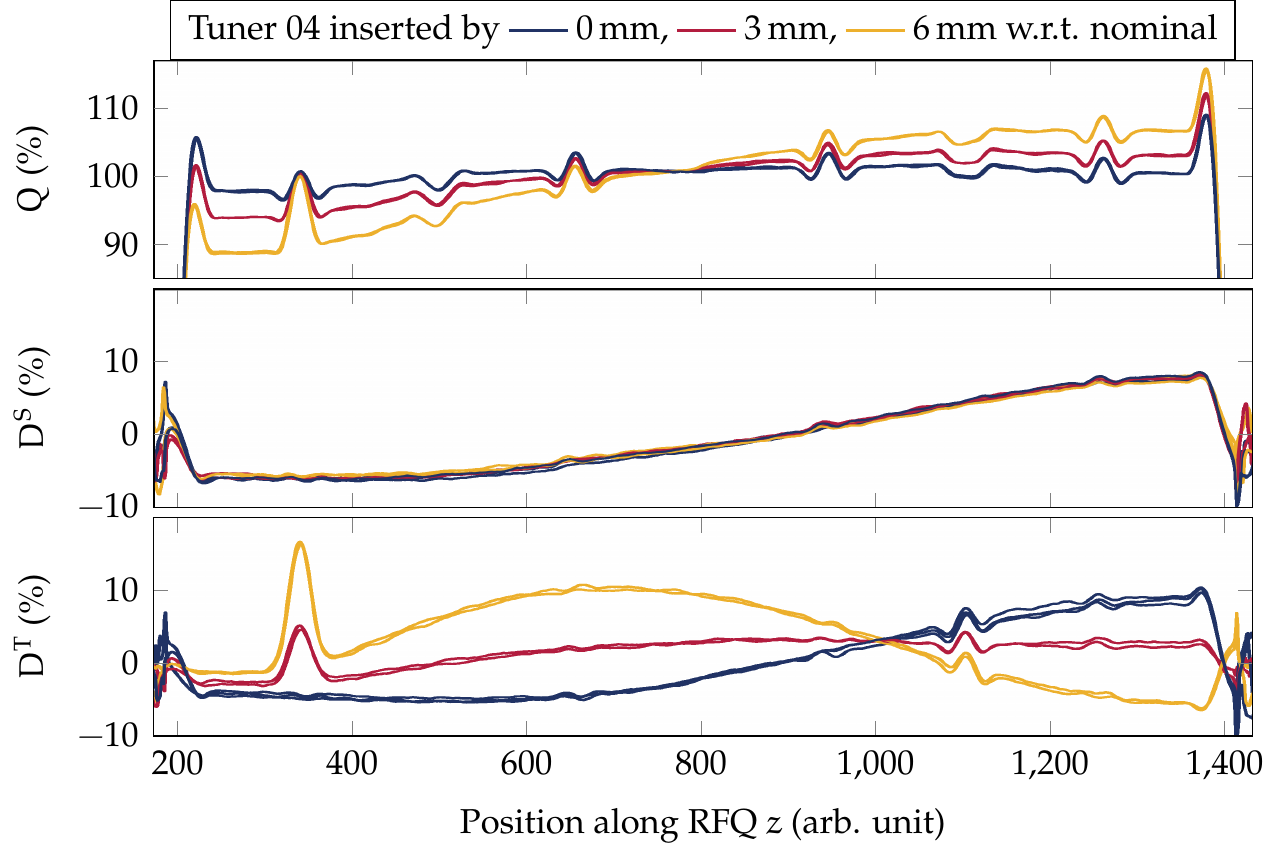}
	\caption{Reproducibility of the field components after movement of tuner~4 (exemplary). The tuner was repeatedly inserted by \SI{0}{\milli\meter} (nominal position), \SI{3}{\milli\meter}, \SI{6}{\milli\meter}, and retracted again.}
	\label{fig:field_tuner_reproducibility}
\end{figure}

A second study was performed to assess errors that originate from tuner movements, i.e. the measurement reproducibility. Several tuners were moved from \SI{0}{\milli\meter} (nominal position in perfect RFQ) to \SI{3}{\milli\meter}, \SI{6}{\milli\meter}, and back to nominal, after which the sequence was repeated. Results for tuner~4 are shown in Fig.~\ref{fig:field_tuner_reproducibility} (without any averaging as discussed above). The error observed in this study is virtually the same as the one observed during the repeatability study (Fig.~\ref{fig:field_error_averaging}), indicating that the overall measurement error was dominated by the bead-pull setup itself, while mechanical hysteresis played a negligible role. 

From Fig.~\ref{fig:field_tuner_reproducibility}, an estimate of the influence of a single tuner on the field distribution could be obtained. An insertion by \SI{3}{\milli\meter} reduced the $Q$ component by roughly \SI{3}{\percent} in the vicinity of the tuner, and modifies $D^\text{T}$ by roughly \SI{6}{\percent}. Contrarily, it has almost no influence on $D^\text{S}$---as one would expect because of the dipole component definitions, each incorporating two opposite quadrants~[Eq.~\eqref{eq:bpull_quad_dip_components}]. In the bottom plot at $z\approx 700$ it becomes clear that the relationship is nonlinear; the influence of a tuner increases with its insertion depth.


\subsection{Augmented tuning algorithm}

The PIXE-RFQ was tuned with an improved version of the algorithm developed for the HF-RFQ by Koubek~\cite{Koubek2017rf, Koubek2017rfrep}, which is based on the tuning scheme for four-vane RFQs described in Ref.~\cite{Wangler2008RF}. For the PIXE-RFQ, the response matrix was augmented by the authors to also include the frequency, such that it could be tuned at the same time as the field. 

\subsubsection{Previous tuning algorithm used for HF-RFQ}

Koubek~\cite{Koubek2017rf} defined (in a slightly different notation)
\begin{equation}
\vec{x}_\text{cur} = 
\begin{bmatrix}
Q_1 & \cdots & Q_n & D^\text{S}_1 & \cdots & D^\text{S}_n & D^\text{T}_1 & \cdots & D^\text{T}_n
\end{bmatrix}^\T \in\mathbb{R}^{3n}
\end{equation}
as the vector of quadrupole and dipole field amplitudes $Q_i$, $D^\text{S}_i$, $D^\text{T}_i$ currently measured in the RFQ at the discrete field sampling points $i=1,\ldots,n$. The corresponding target values are summarized in the vector
\begin{equation}
\vec{x}_\text{trg} = 
\begin{bmatrix}
Q_{\text{trg},1} & \cdots & D^\text{T}_{\text{trg},n}
\end{bmatrix}^\T
\in\mathbb{R}^{3n}.
\end{equation}
Similarly, the current tuner positions~$t_j$ and target tuner positions~$t_{\text{trg},j}$, \mbox{$j=1,\ldots,m$} are collected in
\begin{equation}
\vec{t}_\text{cur} = 
\begin{bmatrix}
t_1\\ 
\vdots\\ 
t_m
\end{bmatrix}\in\mathbb{R}^m,
\quad
\vec{t}_\text{trg} = 
\begin{bmatrix}
t_{\text{trg},1}\\ 
\vdots\\ 
t_{\text{trg},m}
\end{bmatrix}\in\mathbb{R}^m.
\label{eq:tuning_koubek_t}
\end{equation}
By moving each tuner individually by some distance (all other tuners remain in nominal position), the response matrix~$\vec{R}=\partial\vec{x}_\text{cur}/\partial\vec{t}_\text{cur}\in\mathbb{R}^{3n\times m}$ is obtained. Each matrix entry quantifies the effect of one tuner on $Q$, $D^\text{S}$, $D^\text{T}$ at one sampling point in first-order approximation. In a perfectly tuned constant-voltage RFQ the quadrupole component equals unity ($Q_{\text{trg},i}=1\;\forall i$) whereas the dipole components vanish (\mbox{$D^\text{S}_{\text{trg},i}=D^\text{T}_{\text{trg},i}=0\;\forall i$}) at all sampling points. Thus, the arising system of equations reads
\begin{equation}
	\underbracex{\vec{x}_\text{trg}-\vec{x}_\text{cur}}{\Delta\vec{x}} = \underbracex{\left(\partial\vec{x}_\text{cur}/\partial\vec{t}_\text{cur}\right)}{\vec{R}} \cdot \underbracex{\left(\vec{t}_\text{trg}-\vec{t}_\text{cur}\right)}{\Delta\vec{t}}.
\end{equation}
The new tuner positions required to correct the field distortion are obtained as
\begin{equation}
	\vec{t}_\text{trg} = \vec{t}_\text{cur} + \Delta\vec{t} = \vec{t}_\text{cur} + \vec{R}^\dagger\Delta\vec{x},
\end{equation}
where $\vec{R}^\dagger$ identifies the pseudo-inverse of~$\vec{R}$.

\subsubsection{Including the frequency}

In this work, we augmented the presented system of equations to include also the frequency. The measured frequency is introduced as a dimensionless quantity~$\bar{f}=w_ff$ such that it can be combined with the likewise dimensionless measured field amplitudes $Q_i, D^\text{S}_i, D^\text{T}_i$ by appending it to~$\vec{x}_\text{cur}$. Analogously, the target frequency $\bar{f}_\text{trg}=w_ff_\text{trg}$ is appended to~$\vec{x}_\text{trg}$. The normalizing weight~$w_f$ (in units of~\si{\per\hertz}) can be used to control the influence of the frequency compared to the field amplitudes. More precisely, $w_f$ determines the contribution of the frequency deviation~$f_\text{trg}-f$ to the residual $||\Delta\vec{x}-\vec{R}\Delta\vec{t}||$, which is minimized when solving the over-determined system by means of the least-squares method. Larger~$w_f$ translate to higher importance. The frequency is incorporated by appending a new row to~$\vec{R}$:
\begin{equation}
\left[\vec{R}\right]_{3n+1} = \frac{\partial\bar{f}}{\partial\vec{t}_\text{cur}} =
\begin{bmatrix}
	\dfrac{\partial\bar{f}}{\partial t_1} & \cdots & \dfrac{\partial\bar{f}}{\partial t_m}
\end{bmatrix}.
\end{equation}

The algorithm presented in Ref.~\cite{Koubek2017rf} forces the normalized quadrupole component to equal unity, $Q_{\text{trg},i}=1\forall i$, which might not lead to an optimum solution when including the frequency. Instead, we use the relaxed condition \mbox{$Q_{\text{trg},i} = \hat{Q}_\text{trg} \forall i$}, which only requires all quadrupole component samples to be equal to some value~$\hat{Q}_\text{trg}$, not necessarily unity, but close. As an unknown quantity, $\hat{Q}_\text{trg}$ must be brought to the right-hand side. This is accomplished by normalizing the tuner positions by a weight $w_t$, i.e. replacing the $t_i$~by~$\bar{t}_i=w_tt_i$, and $t_{\text{trg},i}$~by~$\bar{t}_{\text{trg},i}=w_tt_{\text{trg},i}$. $\vec{R}$ is augmented with a corresponding new column, and the system
\begin{equation}
\let\oldarraystretch\arraystretch
\renewcommand{\arraystretch}{1.35}
\underbracex{\begin{bmatrix}
\xq  -Q_1\\ 
\xq  \vdots\\ 
\xq  -Q_n\\ 
\xs   -D^\text{S}_1\\ 
\xs   \vdots\\ 
\xs   -D^\text{S}_n\\ 
\xt -D^\text{T}_1\\ 
\xt \vdots\\ 
\xt -D^\text{T}_n\\
\Delta\bar{f}
\end{bmatrix}}{\Delta\vec{x}}
=
\underbracex{\begin{bmatrix}
\xq \frac{\partial Q_1}{\partial \bar{t}_1} & \xq \frac{\partial Q_1}{\partial \bar{t}_2} & \xq \cdots & \xq\frac{\partial Q_1}{\partial \bar{t}_m} & \xq -1\\
\xq \vdots & \xq \ddots & \xq \ddots & \xq \vdots & \xq \vdots \\
\xq \frac{\partial Q_n}{\partial \bar{t}_1} & \xq \frac{\partial Q_n}{\partial \bar{t}_2} & \xq \cdots & \xq \frac{\partial Q_n}{\partial \bar{t}_m} & \xq -1\\
\xs \frac{\partial D^\text{S}_1}{\partial \bar{t}_1} & \xs \frac{\partial D^\text{S}_1}{\partial \bar{t}_2} & \xs \cdots & \xs \frac{\partial D^\text{S}_1}{\partial \bar{t}_m} & \xs 0\\
\xs \vdots & \xs \ddots & \xs \ddots & \xs \vdots & \xs \vdots\\
\xs \frac{\partial D^\text{S}_n}{\partial \bar{t}_1} & \xs \frac{\partial D^\text{S}_n}{\partial \bar{t}_2} & \xs \cdots & \xs \frac{\partial D^\text{S}_n}{\partial \bar{t}_m} & \xs 0\\
\xt \frac{\partial D^\text{T}_1}{\partial \bar{t}_1} & \xt \frac{\partial D^\text{T}_1}{\partial \bar{t}_2} & \xt \cdots & \xt \frac{\partial D^\text{T}_1}{\partial \bar{t}_m} & \xt 0\\
\xt \vdots & \xt \ddots & \xt \ddots & \xt \vdots & \xt \vdots \\
\xt \frac{\partial D^\text{T}_n}{\partial \bar{t}_1} & \xt \frac{\partial D^\text{T}_n}{\partial \bar{t}_2} & \xt \cdots & \xt \frac{\partial D^\text{T}_n}{\partial \bar{t}_m} & \xt 0\\
\frac{\partial\bar{f}}{\partial \bar{t}_1} & \frac{\partial\bar{f}}{\partial \bar{t}_2} & \cdots & \frac{\partial\bar{f}}{\partial \bar{t}_m} & 0
\end{bmatrix}}{\vec{R}}
\underbracex{\begin{bmatrix}
\bar{t}_{\text{trg},1}-\bar{t}_1\\ 
\vdots\\ 
\bar{t}_{\text{trg},m}-\bar{t}_m\\
\hat{Q}_\text{trg}
\end{bmatrix}}{\Delta\vec{t}},
\label{eq:tuning_system}
\renewcommand{\arraystretch}{\oldarraystretch}
\end{equation}
emerges, where $\Delta\bar{f}=\bar{f}_\text{trg}-\bar{f}=w_f\Delta f$, and all quantities are dimensionless.


\subsubsection{Matrix inversion by SVD}

Eq.~\eqref{eq:tuning_system} must be solved to obtain the tuner corrections. One possible solution is given by
\begin{equation}
	\Delta\vec{t}=\vec{R}^\dagger\Delta\vec{x},\label{eq:tuning_moore_penrose_inverse}
\end{equation}
where $\vec{R}^\dagger=(\vec{R}^\T\vec{R})^{-1}\vec{R}$ denotes the Moore-Penrose inverse~\cite{Bronshtein2015Handbook} (pseudo-inverse) of the generally non-square~$\vec{R}$. Koubek~\cite{Koubek2017rf,Koubek2017rfrep} pointed out that $\vec{R}$ is potentially ill-conditioned and proposed to use a special method based on singular value decomposition~(SVD)~\cite{Bronshtein2015Handbook} to compute the inverse. 

To simplify the notation, we identify~$N=3n+1$ and~$M=m+1$ corresponding to a setup with $n$~longitudinal field sampling points and $m$~tuners. The SVD of~$\vec{R}\in\mathbb{R}^{N\times M}$ is given as
\begin{equation}
    \vec{R}=\vec{U}\vec{\Sigma}\vec{V}^\T,
\end{equation}
where $\vec{U}\in\mathbb{R}^{N\times N}$, $\vec{V}\in\mathbb{R}^{M\times M}$ are orthonormal matrices, and $\vec{\Sigma}\in\mathbb{R}^{N\times M}$ is a rectangular diagonal matrix
whose diagonal entries~$\sigma_1,\ldots,\sigma_M$ are the singular values of $\vec{R}$ in descending order. Note that $M\leq N$ is an essential requirement of the algorithm, which can always be achieved by increasing the number of sampling points. The pseudo-inverse can then be constructed as \mbox{$\vec{R}^\dagger=\vec{V}\vec{\Sigma}^\dagger\vec{U}^\T$},
where $\vec{\Sigma}^\dagger$ is obtained by inverting each diagonal entry~$\sigma_i$ of $\vec{\Sigma}$ and transposing the result.

For ill-conditioned, almost singular $\vec{R}$, the reciprocals~$1/\sigma_i$ of the smaller singular values (larger~$i$) approach infinity. This can lead to invalid solutions, i.e. tuner adjustments that lie outside of the physical tuner movement range. Koubek~\cite{Koubek2017rf, Koubek2017rfrep} proposed to circumvent this problem by consecutively setting the largest~$1/\sigma_i$ of~$\vec{\Sigma}^\dagger$ to zero. We define
\begin{equation}
    \vec{\Sigma}^\dagger_k = \diag\begin{bmatrix}
        \frac{1}{\sigma_1} & \cdots & \frac{1}{\sigma_{M-k}} & 0 & \cdots & 0
    \end{bmatrix} \in\mathbb{R}^{M\times N}
\end{equation}
as the matrix $\vec{\Sigma}^\dagger$ where the $k$ largest entries have been set to zero, with \mbox{$k = 0,\ldots,(M-1)$}. Note that $\vec{\Sigma}^\dagger_0=\vec{\Sigma}^\dagger$ is the initial Moore-Penrose inverse, whereas $k=M$ leads to $\vec{R}^\dagger=\vec{0}$ and no tuner movements at all. The matrices give~$M$ possibly useful solutions
\begin{equation}
	\Delta\vec{t}_k = \vec{V}\vec{\Sigma}^\dagger_k\vec{U}^\T\Delta\vec{x}. \label{eq:tuning_solution_k}
\end{equation}
Naturally, solutions with tuner adjustments outside the physical movement range must be discarded. The remaining solutions are checked by computing the prediction~$\vec{\tilde{x}}_{\text{trg},k}$ for corrected field and frequency using the original response matrix:
\begin{equation}
	\vec{\tilde{x}}_{\text{trg},k} = \vec{x}_\text{cur} + \vec{R}\Delta\vec{t}_k. \label{eq:tuning_prediction_k}
\end{equation}
The tuner movement for the current tuning step is chosen as that $\Delta\vec{t}_k$ whose corresponding $\vec{\tilde{x}}_{\text{trg},k}$ best fulfills the requirements: a quadrupole component as flat as possible, dipole components and frequency deviation as close to zero as possible.

After applying one tuner adjustment, field and frequency are measured again and a new vector $\vec{x}_\text{cur}$ emerges. The process is repeated until the measured field components match the requirements to a desired accuracy~\cite{Wangler2008RF, Koubek2017rf, Koubek2017rfrep}. Koubek showed that it is sufficient to use only the initial $\vec{R}$ and choose a different solution~[Eq.~\eqref{eq:tuning_solution_k}], i.e. a different value for $k$, for each tuning iteration~\cite{Koubek2017rf, Koubek2017rfrep}. This way, time-consuming re-measurement of $\vec{R}$ for each tuning iteration is avoided.


\subsection{Tuning steps}

The tuning procedure can be structured into two phases: at first, the response matrix~$\vec{R}$ was measured by means of individual tuner movements. Then, two corrective tuner movements were carried out.

\subsubsection{Measurement of response matrix}

The entries of~$\vec{R}$ from Eq.~\eqref{eq:tuning_system} were determined by means of spectrum and bead-pull measurements. Each matrix column was obtained by moving the corresponding tuner a certain length while all other tuners remained in nominal position. The probing tuner movements~$\Delta t_j$ should resemble the anticipated corrective movement as close as possible as $\vec{R}$ contains only first-order approximations of the de facto nonlinear responses: \mbox{$\partial X_i/\partial t_j\approxeq\Delta X_i/\Delta t_j=\text{const.}$}, where \mbox{$X\in\{Q, D^\text{S}, D^\text{T}\}$}. From Fig.~\ref{fig:field_tuner_reproducibility} it can be seen that a movement of \SI{3}{\milli\meter} affects $D^\text{T}$ to an amount similar to the detuning in $D^\text{T}$. Therefore, the derivatives in~$\vec{R}$ were approximated as difference quotients where $\Delta t_j=\SI{3}{\milli\meter}$, i.e. each tuner was retracted by \SI{3}{\milli\meter}. Naturally, the tuner normalization constant was chosen as $w_t=(\SI{3}{\milli\meter})^{-1}$.

\begin{figure}[tbh]
    \centering
    
    \subfloat[]{%
        \includegraphics[width=0.8\linewidth]{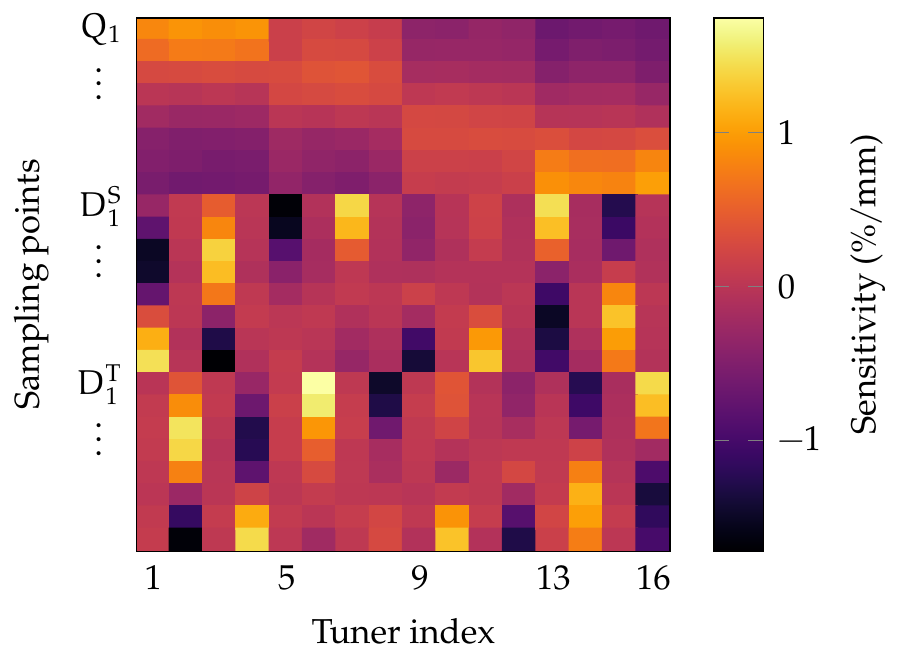}%
        \label{fig:response_matrix}%
    }
    
    \subfloat[]{%
        \includegraphics[width=0.8\linewidth]{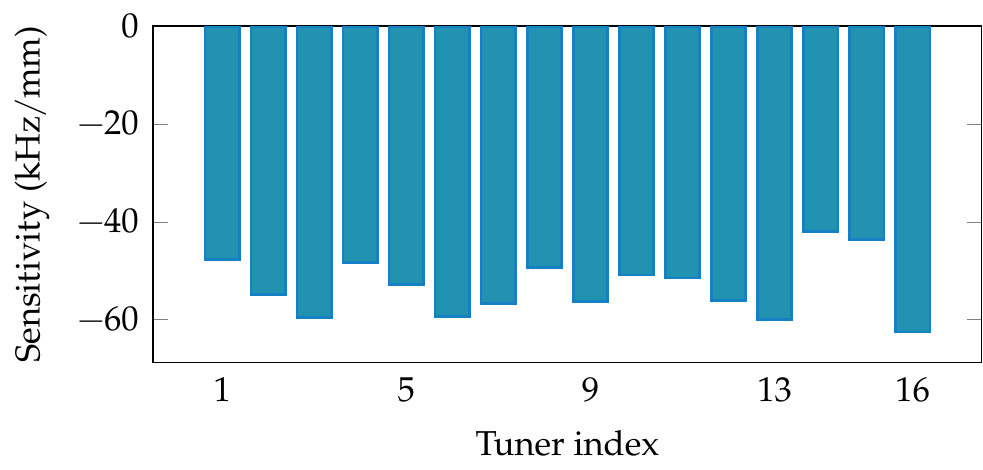}%
        \label{fig:response_frequency}%
    }

\caption{Visual representation of response matrix~$\vec{R}$ except for the last row and column~\protect\subref{fig:response_matrix}. Each entry represents the sensitivity of the $Q$ (upper third), $D^\text{S}$ (middle third), or $D^\text{T}$ component (lower third) at one sampling point with respect to the tuner movement given in units of~\si{\percent\per\milli\meter} (without normalization). The responses of the TE\textsubscript{210} eigenfrequency to the movement of each tuner, i.e. the last row of~$\vec{R}$, is shown in \protect\subref{fig:response_frequency}. A negative derivative means that the frequency is lowered when the tuner is retracted, as the total cavity inductance is increased.}
\end{figure}

In Fig.~\ref{fig:response_matrix}, a visual representation of the response matrix is shown. The last row and column are omitted here in order to show only the field sensitivities. All four tuners at the same longitudinal position (e.g. tuners~1 to~4) have approximately the same effect on $Q$, where tuners located at the extremities of the RFQ induce slightly stronger field tilts than those near the center. Each tuner has a strong effect on the dipole component comprising the quadrant in which the tuner is located, while the other dipole component is influenced only marginally~[Eq.~\eqref{eq:bpull_quad_dip_components}]. Two tuners located opposite of each other (e.g. tuners 1~and~3, or 2~and~4, see Fig.~\ref{fig:tuner_labels}) have opposite effects on the respective dipole component. The frequency---an integral quantity proportional to the total field energy---is affected by each tuner with roughly the same sensitivity of \SI{-50}{\kilo\hertz\per\milli\meter}~[Fig.~\ref{fig:response_frequency}]. Individual deviations are attributed to the inhomogeneous capacitance and inductance distributions caused by vane modulation, misalignments, and design choices, which in fact motivate the RFQ tuning.

In each tuning step, the frequency was enforced to equal the nominal frequency of \SI{749.480}{\mega\hertz} to an accuracy ensuring that no significant frequency error would be introduced by the tuning algorithm itself. We selected~$w_f=(10^4\,\si{\hertz})^{-1}$ such that the predicted frequency obtained from $\vec{\tilde{x}}_{\text{trg},k}$~[Eq.~\eqref{eq:tuning_prediction_k}] matched the desired frequency with an error smaller than \SI{0.1}{\kilo\hertz}, while still keeping the weight as small as possible.

\subsubsection{Corrective tuner movements}

\begin{figure}[tbh]
	\centering
	
	\subfloat[]{
	    \includegraphics[width=\linewidth]{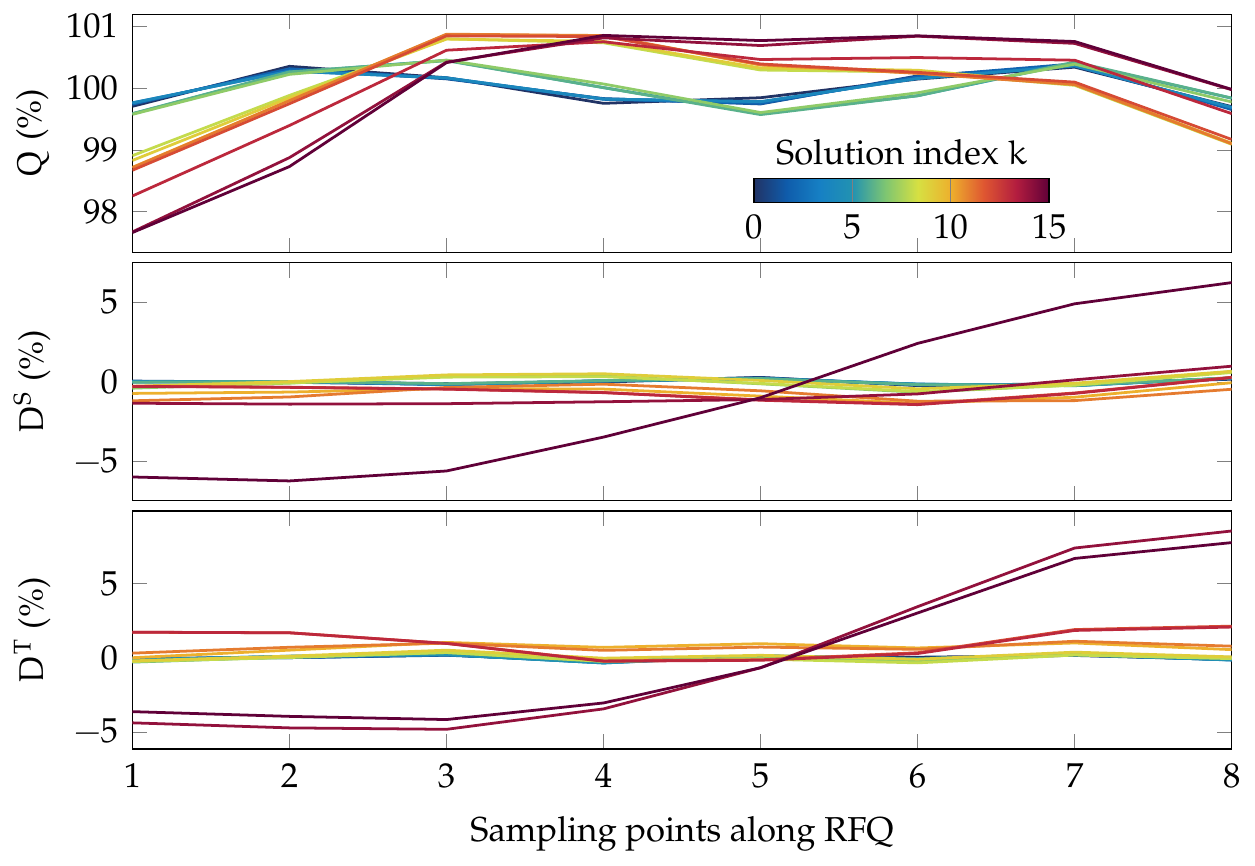}%
	    \label{fig:tuning_step1_predicted_fields}%
	}
	
	\subfloat[]{
	    \includegraphics[width=\linewidth]{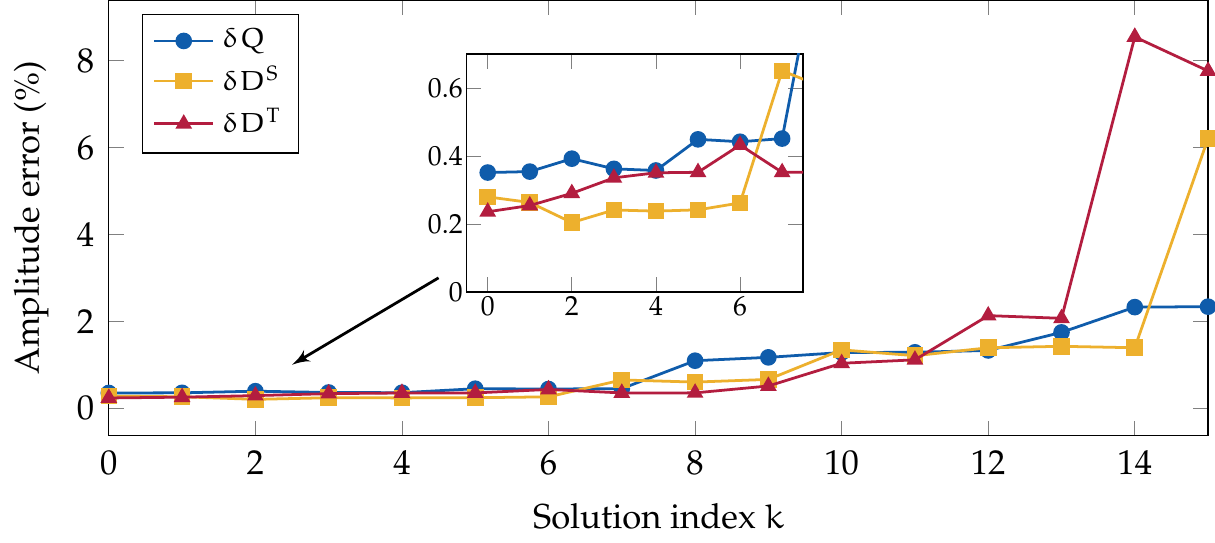}%
	    \label{fig:tuning_step1_predicted_errors}%
	}

	\caption{Predicted field components~\protect\subref{fig:tuning_step1_predicted_fields} after applying the tuner adjustments yielded during tuning step~1 by discarding the smallest~$k$ singular values from the response matrix. Corresponding field component errors are reported in~\protect\subref{fig:tuning_step1_predicted_errors}.}
	\label{fig:tuning_step1_predicted}
\end{figure}

With Eq.~\eqref{eq:tuning_prediction_k}, $M$ possible solutions $\Delta\vec{t}_k$~[Eq.~\eqref{eq:tuning_solution_k}] yield $M$ predictions $\vec{\tilde{x}}_{\text{trg},k}$ of the field distribution after the tuning step, shown in Fig.~\ref{fig:tuning_step1_predicted_fields} for the first tuning step. Corresponding amplitude errors are reported in Fig.~\ref{fig:tuning_step1_predicted_errors}. Solutions with lower~$k$ (less singular values eliminated) generally deliver better corrections. In Refs.~\cite{Koubek2017rf,Koubek2017rfrep}, Koubek reported that---with the original version of the algorithm---many of the possible solutions with small~$k$ yielded tuner adjustments that were outside the physical tuner movement range. In the present case this could not be confirmed; all $k=1,\ldots,M-1$ solutions provide physically possible tuner movements. This is attributed to the fact that the augmented version of the algorithm~[Eq.~\eqref{eq:tuning_system}] includes the frequency, which would strongly disagree with the desired value for unphysically large tuner movements and thus acts in a dampening manner. Furthermore, the response matrix of the HF-RFQ was more affected by measurement noise, as no averaging was performed~\cite{Koubek2017rf,Koubek2017rfrep}. Nevertheless, it is still advantageous to make use of the truncated SVD in the improved algorithm and select a solution ``by eye,'' as solution~$k$ does not strictly provide a better correction than solution~$k+1$.

\begin{figure}[tbh]
	\centering
	\includegraphics[width=\linewidth]{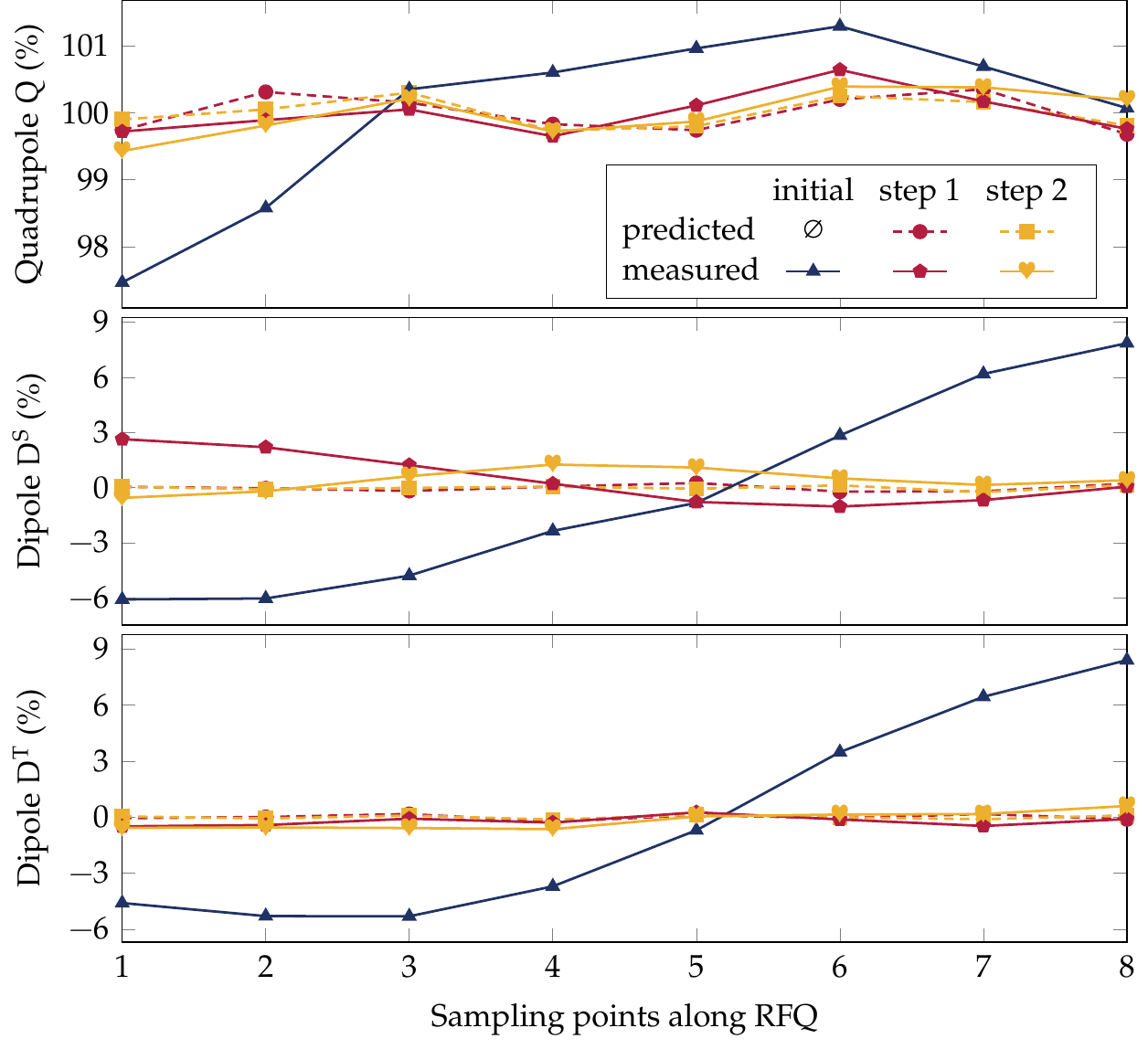}
	\caption{Field components for nominal tuner position (initial) and after the first and second tuning step. Dashed lines indicate the field predicted by Eq.~\eqref{eq:tuning_prediction_k}, while solid lines report the actually measured field.}
	\label{fig:tuning_steps_field}
\end{figure}

\begin{figure}[tbh]
	\centering
	\includegraphics[width=0.8\linewidth]{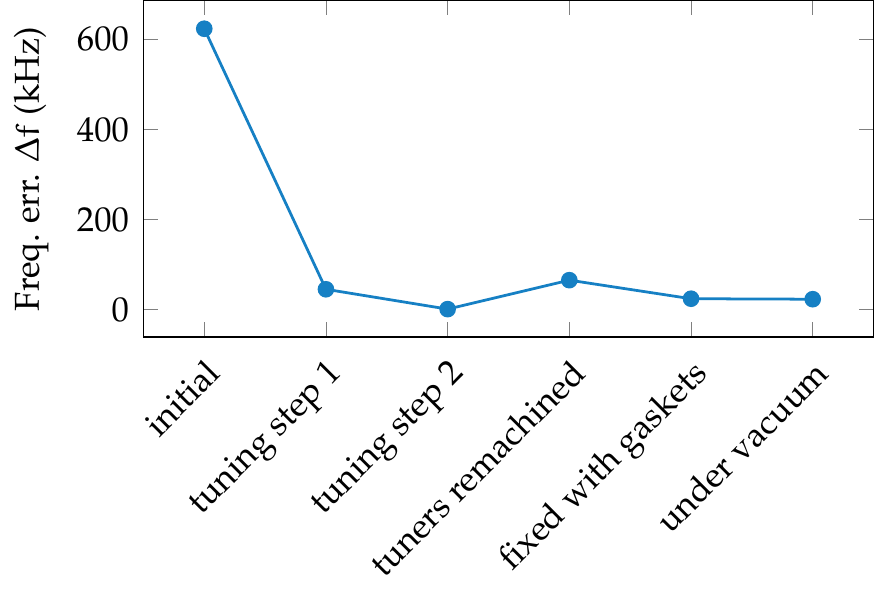}
	\caption{Frequency of the TE\textsubscript{210} eigenmode for nominal tuner position (initial), after the first and second tuning step, and after remachining measured under air and vacuum. The final frequency error amounts to \mbox{$\Delta f=\SI{23.5}{\kilo\hertz}$}.}
	\label{fig:tuning_steps_frequency}
\end{figure}

The PIXE-RFQ was tuned in only two steps after the measurement of~$\vec{R}$. For the first step, $k=1$ was chosen as the option that provides very small errors for all three components while equally correcting $D^\text{S}$,~$D^\text{T}$. The corresponding field prediction is shown in Fig.~\ref{fig:tuning_steps_field} (red dashed line). After applying the tuner adjustments, the frequency error improved from~\SI{624}{\kilo\hertz} to~\SI{46}{\kilo\hertz}~(Fig.~\ref{fig:tuning_steps_frequency}). A significant deviation between field prediction and measurement (red solid line) was observed, in particular for $D^\text{S}$, where an error of nearly~$\pm\SI{3}{\percent}$ remained. The reason is found in the nonlinear relation between field and tuner position, that is represented in the response matrix only by linear approximation.

A second tuning step with the same response matrix~$\vec{R}$ was carried out, however, this time we chose $k=3$. Predicted and measured fields after the second tuning step are reported in Fig.~\ref{fig:tuning_steps_field} (yellow lines). $D^\text{S}$~could be suppressed to an error of \SI{1.3}{\percent}, and the frequency deviation was improved to \SI{1.5}{\kilo\hertz}. This error is of the same order of magnitude as the error arising from the thermometer precision~($\pm\SI{2.6}{\kilo\hertz}$). The second tuning step already saw a slight worsening in~$D^\text{T}$. This indicates that an accuracy limit given either by the overall noise level or the linear approximation was reached. The errors in frequency and field after the second tuning step fulfilled the requirements listed in Section~\ref{sec:tuning_goals}: $\delta Q=\pm\SI{0.6}{\percent}$, $\delta D^\text{S}=\pm\SI{1.3}{\percent}$, and $\delta D^\text{T}=\pm\SI{0.6}{\percent}$. As a third iteration delivered no satisfying predictions of improvement, it was decided to stop the tuning procedure after two steps.

\subsubsection{Tuner recutting}

\begin{figure}[tbh]
	\centering
	\includegraphics[width=0.8\linewidth]{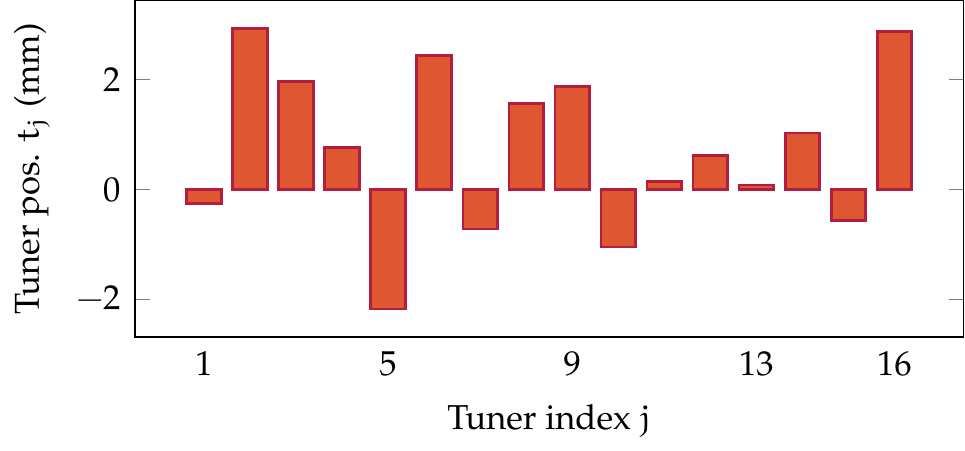}
	\caption{Tuner positions after the final tuning step. A positive (negative) value indicates that the tuner was retracted (inserted) with respect to the nominal position.}
	\label{fig:tuner_positions_final}
\end{figure}

Final tuner positions are reported in Fig.~\ref{fig:tuner_positions_final}. Note that on average the tuners were retracted, as the initial measured frequency was too high and the total cavity inductance had to be increased. All adjustments are considerably smaller than the maximum foreseen movement range of $\pm\SI{11}{\milli\meter}$. The final tuner lengths were determined both from the scale on the tuner tooling~[Fig.~\ref{fig:photo_tuner_tooling_assembly}], which was used during the tuning process itself, and from an additional measurement using a caliper. The errors between the two measurements read up to \SI{60}{\micro\meter}, likely originating in small inclinations of the tuners within the guidance tubes. The average of both values was used for remachining.

\begin{figure}[tbh]
	\centering
	\includegraphics[width=\linewidth]{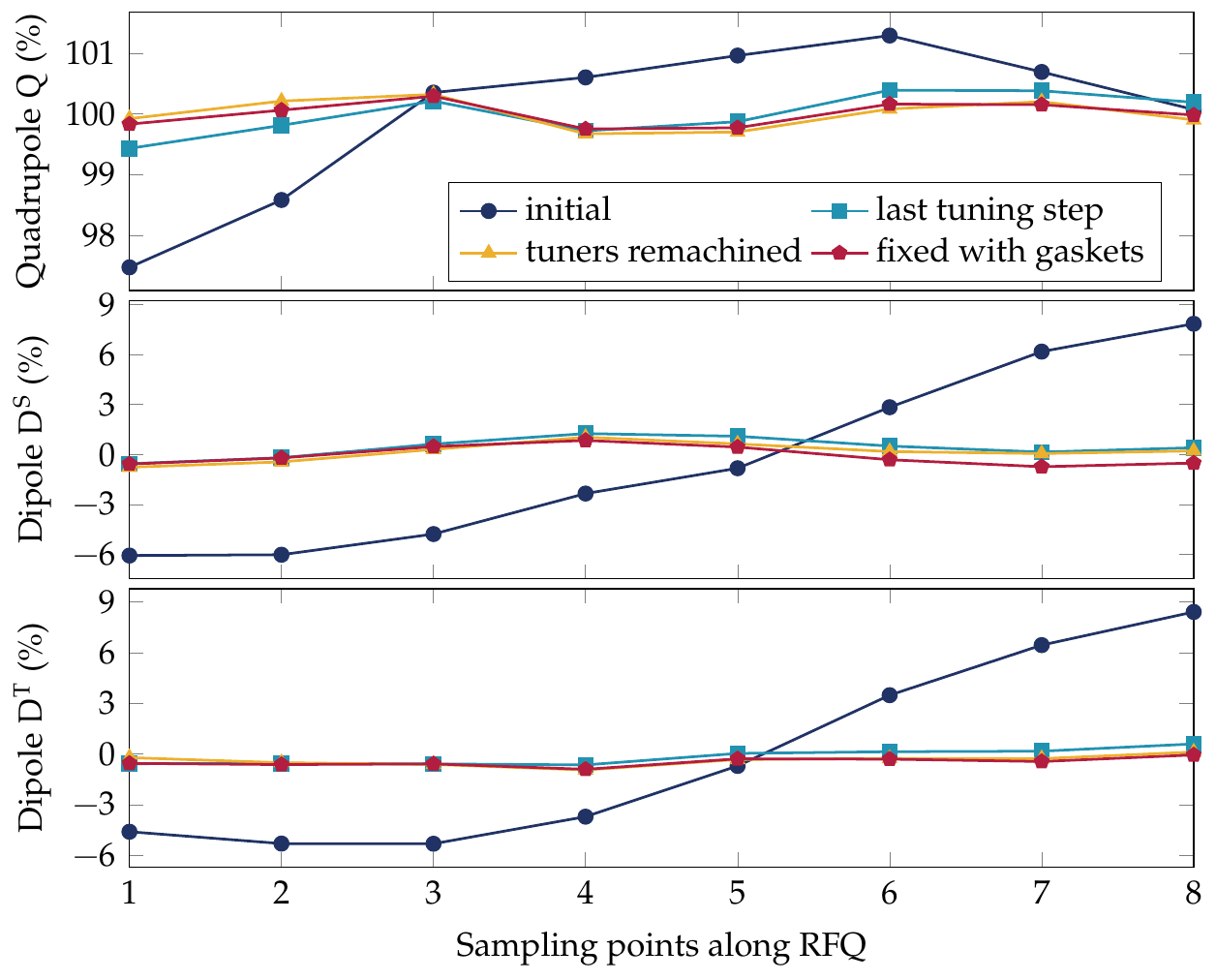}
	\caption{Measured field components before tuning, after tuning, with remachined tuners, and for the final installation of the tuners with copper gaskets.}
	\label{fig:field_before_after}
\end{figure}

Following remachining, a bead-pull measurement was carried out with the tuners fixed to the corresponding flanges without seals as a mean of validation, after which they were installed with vacuum gaskets and a final measurement was performed. The results are reported in Fig.~\ref{fig:field_before_after} in comparison to the initial field and the field after the second tuning step. Corresponding errors are summarized in Table~\ref{tab:tuning_steps}. A small deviation, larger than measurement noise, was observed between the field after the second tuning step and the field after they were remachined and fixed. This deviation likely originates in the finite accuracy of length measurement and material cutting. 

The final frequency measured under vacuum was \SI{23.5}{\kilo\hertz} above the target value. This error was achieved measuring under air, solely correcting for temperature and constant air permittivity (STP value). The deviation can be explained by the fact that response matrix and initial frequency were measured on a rainy day ($\approx\SI{100}{\percent}$ rel. humidity), while the tuning steps were performed under sunny weather ($\approx\SI{50}{\percent}$ rel. humidity). It is likely that the remaining error could have been significantly reduced by correcting for the measured humidity or using a dry flooding gas. Nevertheless, the final frequency error is by more than a factor of two smaller than the tuning range given by a typical water cooling system~($\pm\SI{60}{\kilo\hertz}$).

\begin{table}[tbh]
	\centering
	\caption{Errors in field components and frequency during tuning and subsequent measurements in chronological order.}
	\begin{tabular}{lSSSS}
		\hline\hline
		&&&&\\[-2ex]
		Step & {$\delta Q$ (\si{\percent})} & {$\delta D^\text{S}$ (\si{\percent})} & {$\delta D^\text{T}$ (\si{\percent})} & {$\Delta f$ (\si{\kilo\hertz})} \\
		\hline
		Initial & 2.5 & 7.9 & 8.4 & 623.6 \\
		Tuning step 1 & 0.6 & 2.6 & 0.5 & 45.5\\
		Tuning step 2 & 0.6 & 1.3  & 0.6 & 1.5\\
		Tuners recut & 0.3 & 1.0 & 0.9 & 65.7 \\
		Fixed with gaskets & 0.3 & 0.9 & 0.9 & 24.5 \\
		Under vacuum & {---} & {---} & {---} & 23.5 \\
		\hline\hline
	\end{tabular}
	\label{tab:tuning_steps}
\end{table}

	
	

\section{Quality factor measurement}
\label{sec:Q}

After tuning, the quality factors of the PIXE-RFQ were measured under vacuum to validate the design and provide information on the input power reflection originating in potential over- or under-coupling.

The raw measurement data were the complex reflection coefficient~$S_{11}=\Gamma$ measured in a bandwidth of \SI{1}{\mega\hertz} symmetrically around the TE\textsubscript{210} eigenfrequency. The reflection coefficient describes a circle in the complex plane in very good approximation, from which unloaded ($Q_0$), external ($Q_\text{ex}$), and loaded quality factor ($Q_\ell$) can be obtained, where
\begin{equation}
    \frac{1}{Q_\ell} = \frac{1}{Q_0} + \frac{1}{Q_\text{ex}}. \label{eq:Q_factors_relation}
\end{equation}
Additionally, the coupling coefficient
\begin{equation}
    \beta_\text{c} = Q_0/Q_\text{ex} \label{eq:coupling_coefficient}
\end{equation}
emerges. The parameters were extracted by fitting an ideal circle to $\Gamma$ following the method described in Refs.~\cite{Altar1947Q, Kajfez1994Linear, Kajfez2003Random, Goryashko2015High}. Contrarily to e.g. the three-point method~\cite{Kummer1986Grundlagen, Caspers2010RF}, the circle-fitting method performs implicit averaging by solving a heavily over-determined system of equations, and is thus significantly less susceptible to measurement noise. 

The technique is based on the Möbius transformation
\begin{equation}
	\Gamma = \frac{b_1\bar{\omega} + b_2}{b_3\bar{\omega}+1},\label{eq:Q_factor_moebius}
\end{equation}
with~$\bar{\omega}=2(\omega/\omega_0-1)$, which describes a circle in the complex plane characterized by the parameters~$b_1$, $b_2$, $b_3$. For $\bar{\omega}\to\pm\infty$, $\Gamma \to b_1/b_3$ approaches the detuned reflection, whereas $\bar{\omega}=0$ delivers $\Gamma(\Delta\omega=0)=b_2$ as the reflection coefficient of the loaded resonator. Their distance equals the circle diameter $D=|b_1/b_3-b_2|$, and the coupling coefficient can be determined as $\beta_\text{c} = (2/D-1)^{-1}$.
Furthermore, $Q_\ell = \Im{b_3}$, from which the two remaining quality factors can be calculated using Eqs.~\eqref{eq:Q_factors_relation} and~\eqref{eq:coupling_coefficient}~\cite{Kajfez1994Linear}. The VNA measures $\Gamma(\omega_i)$ at $N$ discrete sampling points~$\omega_i$. Thus, Eq.~\eqref{eq:Q_factor_moebius} can be written in matrix form:
\begin{equation}
\underbracex{\begin{bmatrix}
\Gamma(\omega_1) \\ 
\vdots \\ 
\Gamma(\omega_N)
\end{bmatrix} }{\vec{\Gamma}\in\mathbb{C}^N}
=
\underbracex{\begin{bmatrix}
\bar{\omega}_1 & 1 & -\bar{\omega}_1 \Gamma(\omega_1) \\ 
\vdots & \vdots & \vdots \\ 
\bar{\omega}_N & 1 & -\bar{\omega}_N \Gamma(\omega_N)
\end{bmatrix} }{\vec{K}\in\mathbb{C}^{N\times 3}}
\underbracex{\begin{bmatrix}
b_1 \\ 
b_2 \\ 
b_3
\end{bmatrix}}{\vec{b}\in\mathbb{C}^3},
\end{equation}
a heavily overdetermined system of equations with each row representing one measurement point. The curve described by~$\Gamma$ is not a perfect circle---deviations become stronger with increasing distance from $\omega_0$ (or \mbox{$|\bar{\omega}|\gg 0$}). Furthermore, equidistant frequency sampling implies lower density of measurement points around the critical $\omega_0$. Therefore, a weighting matrix~$\vec{W}$ is introduced, which reduces the significance of measurement points further away from $\omega_0$:
\begin{equation}
	w_i = 1-\left|\Gamma(\omega_i)\right|^2,\qquad \vec{W} = \diag\begin{bmatrix}
	w_1 & \cdots & w_N
	\end{bmatrix} \in \mathbb{R}^{N\times N}.
\end{equation}
Then, the three fitting parameters are obtained as
\begin{equation}
	\vec{b} = \left(\vec{W}\vec{K}\right)^\dagger \vec{W} \vec{\Gamma}.
\end{equation}

\begin{figure}[tbh]
	\centering
	\includegraphics[width=6cm]{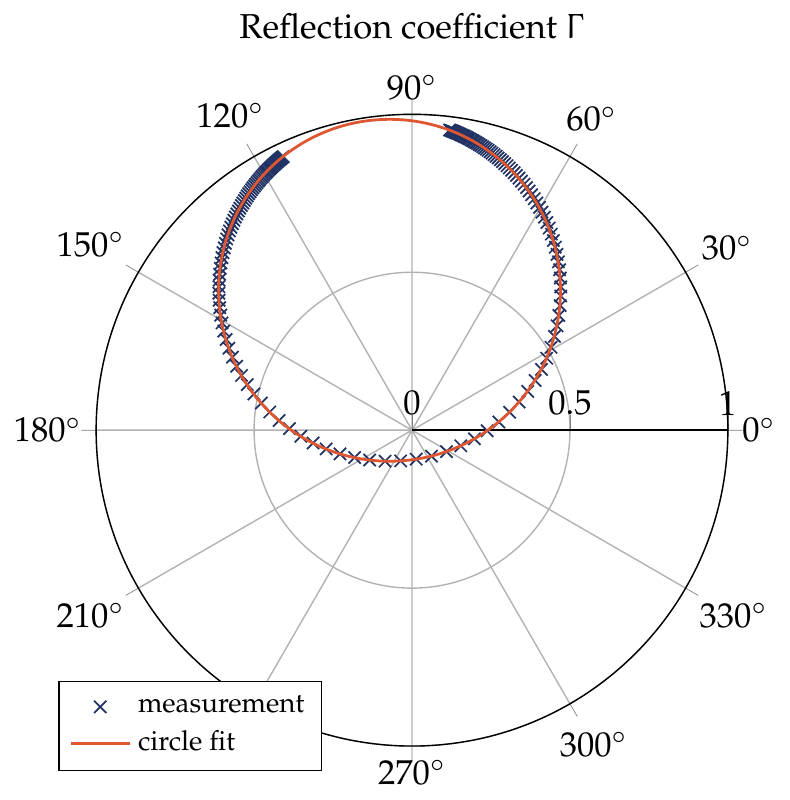}
	\caption{Smith chart of the reflection coefficient~$\Gamma$ measured over a span of \SI{1}{\mega\hertz} and the fitted circle. Only every tenth measurement point is shown for clarity.}
	\label{fig:smith_chart_circle_fitting}
\end{figure}

In Table~\ref{tab:Q_comparison}, measured quality factors and coupling coefficient are compared to the design values~\cite{Pommerenke2019rf}. Very close agreement was observed: the measured~$Q_0$ agrees with the design value with an error smaller than \SI{0.5}{\percent}. This indicates not only the reliability of the power loss calculation carried out at the design stage~\cite{Pommerenke2019rf}, but also the manufacturing quality of the PIXE-RFQ.

\begin{table}[tbh]
	\centering
	\caption{Quality factors measured using the circle-fitting method compared to the design values.}
	\begin{tabular}{lcrrr}
		\hline\hline
		&& Measured & Design & Rel. err.\\
		\hline
		Unloaded quality factor & $Q_0$ &  6018 & 5995 & \SI{0.4}{\percent}\\
		External quality factor & $Q_\text{ex}$ & 5099 & 4796 & \SI{6.3}{\percent}\\
		Loaded quality factor & $Q_\ell$ & 2756 & 2664 & \SI{3.5}{\percent}\\
		Coupling coefficient & $\beta_\text{c}$ & 1.18 & 1.25 & \SI{5.6}{\percent}\\
		\hline\hline
	\end{tabular}
	\label{tab:Q_comparison}
\end{table}

A larger deviation of \SI{6}{\percent} was observed in~$Q_\text{ex}$. This translates to a comparable error in~$\beta_\text{c}$ and an error of \SI{3}{\percent} in~$Q_\ell$. The errors could be introduced by imperfect machining, brazing, and alignment of the physical coupler. A \SI{6}{\percent} error in~$\beta_\text{c}$ could correspond to a mechanical error of \SI{300}{\micro\meter} in the coupling loop size. As the coupler represents a complex geometry composed of multipole components assembled by means of bolts, we cannot exclude such an error. (In contrast, different simulation models generally yield errors in~$Q_\text{ex}$ which are by an order of magnitude smaller.)

All measurements reported in this section were conducted under vacuum (approximately~$10^{-6}\,\si{\milli\bar}$). An interesting observation was made in that the measured coupling increased by approximately \SI{2}{\percent} (from~1.16 to~1.18) when the RFQ cavity was evacuated. A possible explanation is found in the mechanical structure of the power coupler: The inner conductor is only supported by the coupling loop and a polyether ether ketone (PEEK) rf window~\cite{Pommerenke2019rf}. Subjected to the pressure difference between vacuum and exterior atmosphere, the PEEK window could deform and push inwards the inner conductor. This would slightly increase the effective area of the coupling loop, resulting in a larger coupling coefficient.

The discrepancy highlights that it is important to design the coupler with an over-coupling margin of few ten percent. Although the PIXE-RFQ coupler features a rotatable flange that allows for fine-tuning the coupling coefficient by means of measurements, we decided not to take advantage and rather accepted an over-coupling of~\SI{18}{\percent} ($\beta_\text{c}=1.18$). This translates to an input reflection of~$\SI{-22}{\dB}$. If the RFQ cavity consumes $\SI{65.0}{\kilo\watt}$ of peak power in terms of surface losses~\cite{Pommerenke2019rf}, an input power of $\SI{65.5}{\kilo\watt}$ is required, and $\SI{0.5}{\kilo\watt}$, i.e. roughly \SI{0.7}{\percent} of the input power, are reflected back towards the generator. These values are well acceptable for operation.

\section{Summary}

In this paper, the low-power rf measurements and tuning procedure carried out at CERN on the one-meter-long \SI{750}{\mega\hertz} PIXE-RFQ are reported.

Before full assembly, bead-pull measurements were performed on mechanical modules to confirm that no severe errors occured during machining and brazing. The field showed near-perfect agreement with the simulation, whereas the frequency error was equal or smaller than \SI{600}{\kilo\hertz} in both cases. The small observed deviations are well within the capabilities of the tuning system.

The main part of this paper is dedicated to the RFQ tuning procedure. Longitudinal and transverse field tilts were compensated by means of sixteen movable piston tuners. The aim of the tuning was to achieve the most flat possible quadrupole component and vanishing dipole components. Furthermore, the operating mode frequency was tuned to the design frequency of \SI{749.480}{\mega\hertz} under vacuum at \SI{22}{\celsius}.

Reliability measurements were carried out before the actual tuning procedure in order to assess the accuracy limits given by bead-pull measurement and tuner tooling. The bead-pull measurement error could be improved to \SI{0.5}{\percent} by performing each measurement thrice. No significant mechanical hysteresis effects were observed. 

The tuning algorithm used for HF-RFQ~\cite{Koubek2017rf} was augmented for the PIXE-RFQ such that frequency and field could be tuned at the same time. The results suggest that including frequency and corresponding weights also leads to a more stable convergence of the tuning iterations. After measuring the response matrix, the PIXE-RFQ was tuned in only two steps. 
Subsequently, the tuners were recut to their final lengths. The measurement under vacuum revealed a remaining frequency error of~$\SI{23.5}{\kilo\hertz}$, which can be explained with the measurement uncertainty with respect to air humidity. The error lies well within the $\pm\SI{60}{\kilo\hertz}$ tuning range of a typical water cooling system ($\pm\SI{5}{\kelvin}$) and is in any case acceptable since no rf accelerating structures are present downstream of the RFQ.

The three quality factors and the coupling coefficient were measured using a circle-fitting method. Excellent agreement with an error smaller than \SI{0.5}{\percent} was observed in~$Q_0$. $Q_\text{ex}$ was measured to be \SI{6}{\percent} higher than the design value. The discrepancy could be attributed to mechanical manufacturing error in the coupler of roughly~\SI{300}{\micro\meter}. The larger~$Q_\text{ex}$ corresponds to a measured coupling coefficient of~1.18 compared to a design value of~1.25. The measured over-coupling of \SI{18}{\percent} requires an extra \SI{0.5}{\kilo\watt}~(\SI{0.7}{\percent}) of rf peak input power. It was decided to not make use of the possibility of rotating the coupler to closer approach critical coupling.


With the low-power rf measurements completed in mid-2020, the PIXE-RFQ has been transported to INFN, Florence, Italy for high-power rf conditioning and beam measurements. The world's smallest RFQ is expected to commence operation towards the end of~2020.


\begin{acknowledgments}
We wish to thank Sebastien Calvo, Yves Cuvet, Serge Mathot, and the staff of the CERN vacuum brazing workshop for their help in successfully tuning the PIXE-RFQ.

This work has been supported by the CERN Knowledge Transfer Group and the Wolfgang Gentner Programme of the German Federal Ministry for Education and Research (BMBF, grant no. 05E15CHA).
\end{acknowledgments}

\bibliography{references.bib}


\end{document}